\documentclass[11pt]{article}
\usepackage{amsmath,sint,epsfig,cite,graphics}
\usepackage{macros_static}
\usepackage{textcomp}
\usepackage{wrapfig}

\begin{document}

\begin{titlepage}

\begin{flushright}
\vskip 0.7cm
DESY-07-059 \\
SFB/CPP-07-17\\
MIT-CTP 3838\\
\end{flushright}

\vskip 0.35cm
\begin{center}
{\Large\bf 
Non-perturbative renormalization of the chromo-magnetic
operator in Heavy Quark Effective Theory and the B$^*$ -- B 
mass splitting.
\\[0.5ex] 
}
\end{center}
\vskip 0.35cm
\vbox{
\centerline{
\epsfxsize=2.8 true cm
\epsfbox{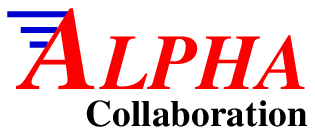}}
}
\vskip 0.1cm
\begin{center}
{
Damiano Guazzini$^{\scriptscriptstyle a}$,
Harvey Meyer$^{\scriptscriptstyle b}$ and  
Rainer Sommer$^{\scriptscriptstyle a}$
}
\vskip 0.5cm
{
$^{\scriptstyle a}$
DESY,
Platanenallee 6, 15738 Zeuthen,  Germany
\vskip 2.0ex
$^{\scriptstyle b}$
Center for Theoretical Physics, Massachusetts Institute of Technology \\
Cambridge, MA 02139, U.S.A.
\vskip 2.0ex
}
\vskip 0.775cm
{\bf Abstract}
\vskip 0.1ex
\end{center}
We carry out the non-perturbative renormalization of the
chromo-magnetic operator in Heavy Quark Effective Theory. At order $1/m$
of the expansion, the operator is responsible for the mass splitting
between the pseudoscalar and vector $B$ mesons. We obtain its 
two-loop anomalous dimension in a Schr\"odinger
functional scheme by successive one-loop conversions to the lattice MS scheme
and the $\overline{\rm MS}$ scheme. We then compute the scale evolution of the
operator non-perturbatively in the $N_{\rm f}=0$ theory between
$\mu\approx0.3$ GeV and $\mu\approx100$ GeV, where contact is made with
perturbation theory. The overall renormalization factor that converts the bare
lattice operator to its renormalization group invariant form is given for the
Wilson gauge action and two standard discretizations of the heavy-quark
action. As an application, we find that this factor brings the previous 
quenched
predictions of the B$^*$ -- B mass splitting closer to the
experimental value than found with a perturbative renormalization. The  same
renormalization factor is applicable to the spin-dependent
potentials of Eichten and Feinberg.

\vskip 0.65cm

\begin{center}
May 2007 
\end{center}

\eject
\vfill
\eject

\end{titlepage}

\section{Introduction \label{s:intro}}

Heavy-light bound states in QCD can be described efficiently
by an expansion in the inverse heavy quark mass.
Already in the early days of the associated effective field theory, 
HQET \cite{stat:eichhill1,hqet:cont3,hqet:cont4},
the mass splitting between
vector and pseudo-scalar  heavy-light mesons served
as a phenomenological argument for the absence of large higher 
order corrections in the expansion. 

Consider QCD with  $\nf$ {\em light} quark flavors and a heavy
flavor, the b-quark, $\nf=4$ being the case realized in Nature.  
The splitting 
\bes
  \Delta m^2 \equiv \mBstar^2-\mB^2  
\ees
then has an asymptotic behavior for large quark mass $\mbeauty$
which is characterized by one renormalization group invariant
(RGI) observable,
\bes
 \label{e:ltwoqcd}
 4 \ltworgi &=& \lim_{\mbeauty\to\infty}  \left\{
 \left[\,2b_0 \gbar^2(\mbeauty)\,\right]^{-\gamma_0/2b_0}
        \Delta m^2 \right\} \,, \\
 && \big(\;\gamma_0=3/(8 \pi^2)\,,\; \;
   b_0=(11-\frac{2}{3}\nf)/(16 \pi^2) \big)\nonumber
\ees
of dimension [mass]$^2$.
Since the limit exists, this quantity is uniquely defined in QCD.
In the above the definition of $\mbeauty$ is irrelevant as long as it
is renormalized at a scale of order $\mbeauty$.

As a rather non-trivial statement, the effective field theory 
predicts that  $\ltworgi$ 
can unambiguously be computed in that theory, where the 
b-quark is treated as static. There it is expressed as
an expectation value 
\bes
  \ltworgi &=& \frac13 \langle B|\Ospinrgi|B\rangle\,/\,\langle B|B\rangle\,, \label{e:lam2}  \\
 \label{e:rgi}
  \Ospinrgi &=& \lim_{\mu\to\infty} 
 \left[\,2b_0 \gbar^2(\mu)\,\right]^{-\gamma_0/2b_0}
                   \Ospin^{S}(\mu)\,,  \label{e:rgiop}
\ees
of $\Ospinrgi$ in the zero-momentum B-meson state $|B\rangle$.
The
operator $\Ospin^{S}$ is related to the bare local operator
\be
   \Ospin(x) = \heavyb(x) {1 \over 2 i} F_{kl}(x) \sigma_{kl} \heavy(x) = 
            \heavyb(x){\boldsymbol\sigma}\!\cdot\!{\bf B}(x)\heavy(x)
            \label{e:ospin}
\ee
by a multiplicative renormalization depending on the adopted scheme 
$S$ and a renormalization scale $\mu$ -- but $\Ospinrgi$ 
neither depends on a scheme nor on a scale. 

The matrix element of the bare operator can be computed non-perturbatively 
by lattice simulations of 
HQET \cite{stat:bbstar,hqet:spinsplitt_gms,hqet:spinsplitt_jlqcd}. 
As stated in these references, a significant source
of uncertainty remained in the connection between the bare operator and
the RGI one (or one renormalized in the $\msbar$ scheme),
which has only been established
perturbatively~\cite{stat:eichhill_1m,stat:ospin_flynn}. 
This uncertainty 
made it impossible to decide whether the splitting is significantly
underestimated in the quenched approximation ($\nf=0$) or not.

In this paper we develop the non-perturbative renormalization
of $\Ospin$. We
follow the general strategy of the 
ALPHA-collaboration~\cite{alpha:sigma,alpha:letter,mbar:pap1}, specialize
it to the operator in question and perform an explicit
computation in the quenched approximation. 
In particular we define a suitable scheme using \SF (SF)
boundary conditions, and compute the 2-loop anomalous dimension in this
scheme. We then evaluate
the scale dependence of the operator non-perturbatively for $\nf=0$,
between $\mu\approx0.3\,\GeV$ and $\mu\approx 100\,\GeV$. Using 
the high energy end of the results and
\eq{e:rgi} supplemented with the 2-loop anomalous dimension,
the connection to the RGI operator is realized.
Finally the total $Z$-factor between bare and
RGI operator is obtained for the Wilson gauge action 
and several HQET discretizations. Readers solely
interested in the final result for the $Z$-factor may find it
in \sect{s:zfact}.

For a comparison to the experimental mass splitting
{\em at finite mass}, it is important to include radiative corrections
beyond the 1-loop ones incorporated in 
\eq{e:ltwoqcd}. We do this in the form
\be
  \Delta m^2 
  = 2 {\mBstar+\mB \over \Mbeauty}\Cspin(\Mbeauty/\Lambda_{\msbar})\ltworgi   
  + \rmO(\minv) \label{e:splitt2}\,,
\ee 
written in terms of RGI's with a function
$\Cspin$ known up to corrections $\rmO(\alpha^2(\mbeauty)) \approx 4\,\%$
and discussed in some detail in \sect{s:rgi}.
Obviously the perturbative uncertainty can be estimated 
more reliably and reduced by a higher order continuum
perturbative computation in QCD.
We note that the final renormalization 
factor $\zmagrgi$ also applies to 
spin-dependent potentials computed in lattice gauge 
theory (see \sect{s:scheme}). 

The reader is not to 
confuse the present approach with the one
of \cite{hqet:pap1,hqet:pap4}, where, through a
non-perturbative matching between QCD and HQET,
also functions such as $\Cspin$ are determined 
non-perturbatively. While in general
the strategy of \cite{hqet:pap1,hqet:pap4} is essential,
the more traditional path is viable here because
$\Ospin$ does not mix with lower dimensional 
operators.

Before entering the discussion of the renormalization of $\Ospin$,
we briefly address the question of the precision that can be expected 
from \eq{e:splitt2}. For this purpose we boldly also
treat $\Delta m^2_\charm = m^2_{\rm D^*} - m^2_{\rm D}$ in HQET. 
So $\nf=3$ in all places. With a $\nf=3$ 
QCD-parameter of $\Lambda_{\msbar}^{(3)}=300(100)\,\MeV$ and with 
$\Mcharm=1.55\,\GeV$, $\Mbeauty=6.69\,\GeV$ 
\cite{mb:steinhkuehn2}
we find
$\Mcharm/\Mbeauty\approx0.23$  
and $\Cspin(\Mcharm/\Lambda_\msbar)/\Cspin(\Mbeauty/\Lambda_\msbar) = 0.94$~.
HQET then relates the splittings
as $\Delta m^2_\charm / \Delta m^2 = 1.41(2)$, where the uncertainty is due
to the generous error in $\Lambda_{\msbar}^{(3)}$.  With the quenched input
values $\Lambda_{\msbar}=238\,\MeV,\,\Mcharm=1.65\,\GeV,\,\Mbeauty=6.76\,\GeV$ 
\cite{mbar:pap1,mbar:charm1,hqet:pap4,lat06:damiano} this ratio changes 
only slightly, namely to $\Delta m^2_\charm / \Delta m^2 \approx 1.44$.
This is to be
compared to $\Delta m^2_\charm / \Delta m^2=1.14$ from experiment. 

Since the  charm mass is only moderately large, 
such a 25\% deviation is not unexpected. 
Scaling this correction to the B-system, we
expect an accuracy of order 5-10\% for the HQET prediction of $\Delta m^2$.
Earlier quenched approximation estimates with perturbative renormalization
found values for $\Delta m^2$ which were lower than the experimental
number by between 50\%
\cite{stat:bbstar,hqet:spinsplitt_gms} and 20\% \cite{hqet:spinsplitt_jlqcd}. 
Renormalizing the same matrix elements non-perturbatively we will find
the difference to experiment significantly reduced in \sect{s:appl}.

\section{HQET and 
         $\ltworgi$ \label{s:rgi}}

\subsection{Lattice action}
We briefly define the effective theory in  
a lattice regularization, using the notation of 
\cite{zastat:pap1,stat:actpaper}. The heavy quark fields
are taken to have 4 components with the constraint
\be
  \label{e_constraint}
  P_{+}\heavy=\heavy\,,\quad \heavyb P_{+}=\heavyb\,\quad
  P_{+}=\frac12(1+\gamma_0)\,.
\ee
With the lattice backward derivative
\be \label{e:D0lat}
D_0^{\rm W} \heavy(x) = {{1}\over{a}} \left[ \heavy(x) -W^{\dagger} (x-a\hat{0},0)
 \heavy (x-a\hat{0}) \right] \; ,
\ee
and the mass counterterm $ \delta m_{\rm W}$,
the static action (i.e. lowest order HQET) is written as
\be
\label{e:stat_action}
S_{\rm h}^{\rm W} =a^4 {{1}\over{1+a\; \delta m_{\rm W} }} \sum_x
\heavyb(x) (D_0^{\rm W} + \delta m_{\rm W} ) \heavy(x) \;.
\ee
Different gauge connections $W$ have been found to be very useful
to improve the statistical precision in numerical 
simulations~\cite{stat:letter,stat:actpaper}. They play a
r\^ole only when we discuss the non-perturbative results. Until then
the reader may think of $W(x,0)$ as the standard timelike link. In fact
that choice defines the original Eichten-Hill action~\cite{stat:eichhill1}. 

\subsection{Conversion functions}
The operator $\Ospin(x)$ in \eq{e:ospin} is given in terms of
the fields entering \eq{e:stat_action} with the normalization 
specified there. The lattice version $\hat{F}_{\mu\nu}$ 
of the gauge field tensor
is defined by the clover leaf term, see e.g. \cite{impr:pap1}.
$\Ospin$ appears as a first order correction
in $\minv$ in HQET and induces the spin splitting.
Usually, the splitting is written in a form different from \eq{e:splitt2}. 
We want to briefly explain why we choose the latter. 

The more common form is  
\bes
  \mBstar^2-\mB^2 &=& 4\, \Cmagmatch(\mbeauty)\,\lambda_2(\mbeauty)   
  + \rmO(\minv) \label{e:splitt3}
  \\
  \lambda_2(\mbeauty) &=&  {1\over 3}\,
  \langle B|\Ospin^\msbar(\mu=\mbeauty)|B\rangle \label{e:splitt4}
\ees 
where $\Ospin^\msbar$ and $\mbeauty$ are renormalized in 
the $\msbar$ scheme. This is arrived at by starting from the
formal expression 
\bes
  \mBstar-\mB &\sim&  \frac23 \frac1\mbeauty \langle B|\Ospin|B\rangle
  \,/\,\langle B|B\rangle
\,. \label{e:splitt_f} 
\ees
One renormalizes $\Ospin$ in the $\msbar$ scheme, identifies the
mass $\mbeauty$ with the (perturbative) pole mass $\mpoleb$ and
defines the remaining factor as a matching coefficient $\Cmagmatch(\mbeauty)$.
Finally one uses $2\mpoleb = \mBstar+\mB + \rmO(\Lambda)$, dropping
the $\rmO(\Lambda)$ correction. 

The matching coefficient 
$\Cmagmatch(\mbeauty) = 1 + C_1 \gbar^2(\mbeauty)+ \ldots$ is 
independent of the particular matrix element. 
Since $\mbeauty \approx 4\GeV$ is reasonably large and
there is no mixing with lower dimensional operators, 
$\Cmagmatch$ can be approximated by perturbation theory. 
It is known including the $C_1 \gbar^2(\mbeauty)$ 
term~\cite{stat:eichhill_1m}.

In the above form, the matrix element and the HQET parameter 
$\lambda_2$ depend on the arbitrary renormalization scheme
($\msbar$). Such a spurious dependence is easily removed by introducing 
RGIs (see e.g. \cite{hqet:pap3}
and Sect.III.3.1.of\cite{nara:rainer}), in particular $\Ospinrgi$, \eq{e:rgiop}.
It is related to the bare operator $\Ospin$ in a particular
lattice regularization via
\bes
 \Ospinrgi &=& \zmagrgi(g_0) \Ospin \,. 
\ees
We now have 
\bes
  \mBstar^2-\mB^2
  &=& 4 \Cmag(\Mbeauty/\Lambda_{\msbar})\ltworgi   
  + \rmO(\minv) \label{e:splitt1}\,
\ees 
with a function $\Cmag(\Mbeauty/\Lambda_{\msbar})$ 
written in terms
of the RGI mass of the b-quark and the QCD $\Lambda$-parameter.
Using existing perturbative computations 
\cite{stat:eichhill_1m,Falk:1991pz,HQET:sigmabI,HQET:sigmabII}
it is easily evaluated by integration
of the RG equations (see e.g. \cite{hqet:pap3}). 

\FIGURE{
\includegraphics*[width=9.5cm]{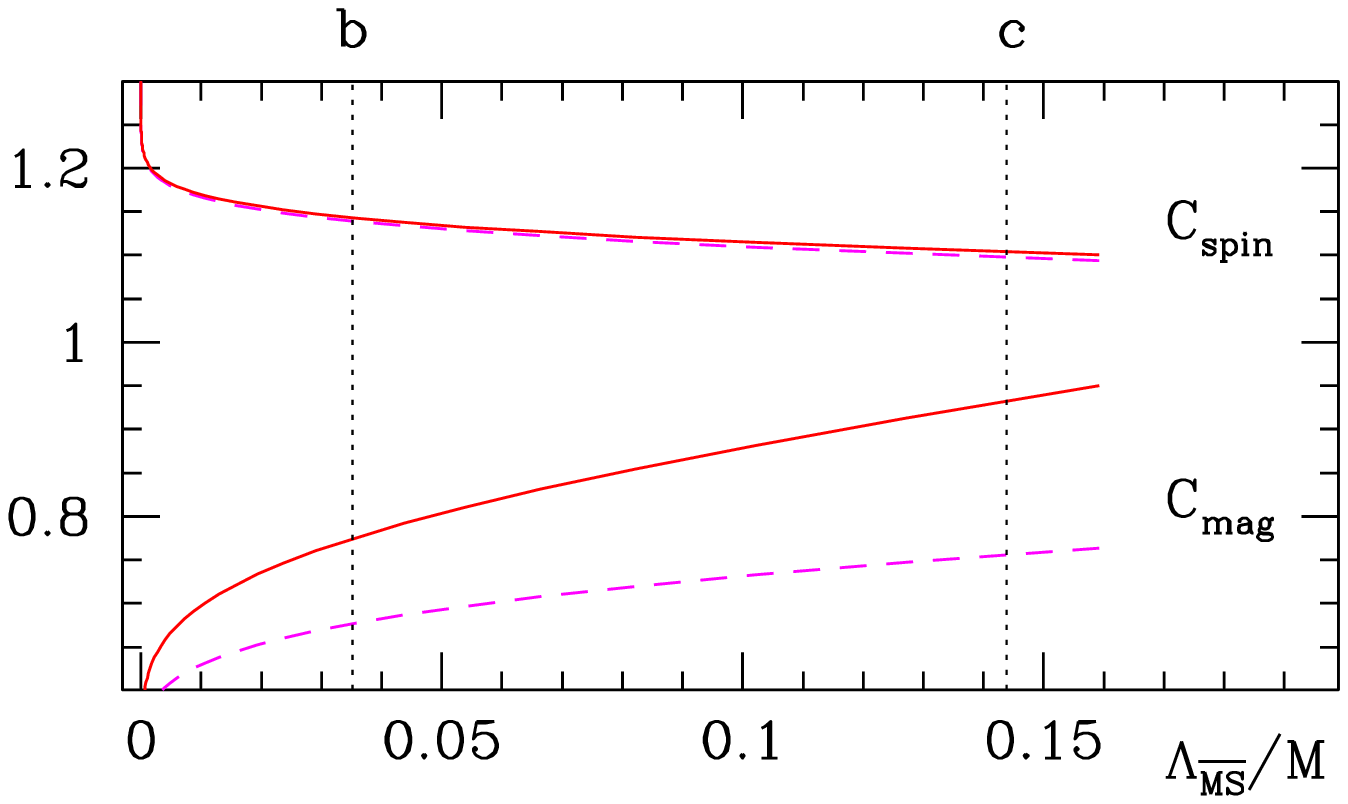}
\caption{\footnotesize
Conversion functions for $\nf=0$. Dashed lines for 1-loop anomalous 
dimension and continuous line for 2-loop $\gamma$. In the latter
case the parametric uncertainty is $\rmO(\alpha^2(\mbeauty))$.
The abscissas of the c quark and the b quark are marked by 
vertical dots. 
}\label{f:convfct}
}

The comparison of the successive perturbative approximations 
for $\Cmag$ in \fig{f:convfct} indicates a relatively large 
perturbative error 
even at the 2-loop level (for the anomalous dimension $\gamma$).
At a scale of $4\,\GeV$ this is somewhat unusual.
However, one can understand this behavior by noting that the 
definition of the 
matching factor $\Cmagmatch$ involves the pole quark mass, which
is unphysical. That mass has a perturbative relation 
to short distance
masses such as $m_{\msbar}$ which is badly behaved (large 
perturbative coefficients). Also the relation pole mass to RGI
mass has this property. 

Our \eq{e:splitt2} follows from choosing directly the RGI mass instead
of the pole mass in \eq{e:splitt_f}. In 
perturbation theory one then has the relation 
\bes
  \Cspin(\Mbeauty/\Lambda_\msbar) &=& {\Mbeauty \over \mpoleb} 
  \Cmag(\Mbeauty/\Lambda_\msbar)\,.
\ees
The function $\Cspin$ is reproduced from \cite{hqet:pap3} in
\fig{f:convfct}. One notices that the two available perturbative
approximations are much closer than they are for $\Cmag$. We expect this 
to hold also at higher orders of perturbation theory. Still,
since the very close agreement of the two approximations for $\Cspin$
may be accidental, we will use an error of $\alpha_s^2(\mbeauty)\approx$4\% at the 
mass of the b-quark. This perturbative error is currently being reduced by an
explicit 3-loop computation \cite{HQET:sigmabIII}.

In summary, the form \eq{e:splitt2} is written in terms of 
properly defined RGI's and  $\Cspin$ appears to have a well 
behaved perturbative 
expansion. This is a good basis for a computation of the mass
splitting.

\section{\SF renormalization scheme \label{s:scheme}}

\subsection{Definition  \label{s:def}}
We want to formulate a renormalization condition
for $\Ospin$ in a finite volume, which allows us to 
carry out a non-perturbative computation of the
associated renormalization factor $\zmagrgi$,
following the general strategy of \cite{mbar:pap1}.
We choose \SF boundary conditions since this allows us
to perform accurate simulations and also perturbative
computations; see \cite{nara:rainer} for a review.
There is an additional reason for this choice. 
With any kind of periodic boundary conditions, any 
correlation function with $\Ospin$ vanishes at tree-level.
In order to avoid this, we choose boundary conditions
which induce a non-trivial background field $F_{\mu\nu}\neq 0$
at tree level. This ensures a good signal in the 
MC simulations at weak coupling and means that a 1-loop 
computation is sufficient to compute the Z-factor 
up to and including
$\rmO(g^2)$. Since the operator does not contain any light
fermion fields, we avoid these altogether in
the definition of the correlation functions. 
For $\nf=0$ we then end up with a pure gauge theory definition
(without valence quarks). Furthermore one easily sees that the
possible dimension six operators which are necessary for
the $\Oa$-improvement of
$\Ospin$ do not contribute here. 
As is well-known, there is also no mixing with 
$ \Okin=\heavyb \vec D^2 \heavy$, the other dimension-5 
operator of HQET.\footnote{The static action is invariant under
the space-dependent transformation 
$
        \delta \heavy(x) = \omega(\vecx)\, \heavy(x)\,,\;
        \delta \heavyb(x) = -\omega(\vecx)\, \heavyb(x) \,, \;
$
which corresponds to the conservation of the local
b-quark number \cite{zastat:pap1}. While 
$\Ospin$ is invariant under this symmetry transformation, 
$\Okin$ is not.} 

These considerations motivate the following choice. We take
an $L_0\times L_1 \times L_2 \times L_3$ geometry and
adopt Dirichlet boundary conditions inducing a 
background field as in \cite{alpha:SU3}. But we choose
Dirichlet conditions in the 3-direction, 
\be\label{e:Ubound}
  U(x,\mu)|_{x_3=0} = \exp(a C)\,, \quad
  U(x,\mu)|_{x_3=L_3} = \exp(a C')\,, \quad \mu=0,1,2 \,,
\ee
keeping periodic boundary conditions with respect to $x_0,x_1,x_2$ 
(remember that time is already distinguished in the static quark action). 
The Abelian fields 
\bes 
   C&=& {i\over L} \diag(\angle_1,\angle_2,\angle_3)= {i\over L}
   \diag(-\pi/3,0,\pi/3)\,,  \nonumber 
  \\[-1.5ex] \label{e:CC} \\[-1.5ex]
   C'&=&  {i\over L} \diag(\angleprime_1,\angleprime_2,\angleprime_3) 
    = {i\over L} \diag(-\pi,\pi/3,2\pi/3)\,, \nonumber
\ees
of ``point A'' \cite{alpha:SU3} are chosen and we set
$L=L_0=L_1=L_2$. The classical solution
then has non-vanishing $F_{\mu 3}$, independent of the space-time
position. Note that
the strength of the fields scales with $L$.

A natural choice of a renormalization condition is then
\be
  \zmagsf(L) {L^2 \langle\, S_1(x+\frac{L}2\hat{0}) \, \Ospin(x) \,\rangle 
      \over
      \langle\, S_1(x+\frac{L}2\hat{0}) \, S_1(x) \,\rangle  }
  =  \left.
     {L^2 \langle\, S_1(x+\frac{L}2\hat{0}) \, \Ospin(x) \,\rangle 
      \over
      \langle\, S_1(x+\frac{L}2\hat{0}) \, S_1(x) \,\rangle  }
     \right|_{g_0=0} \,, \label{e:zspindef} 
\ee
with $x_3=L_3/2=L/2$. The spin operator
\be
  \sk(x) = {{1}\over{1+a\; \delta m_{\rm W} }}
  \heavyb(x)\sigma_k W^{\dagger} (x-a\hat{0},0) \heavy (x-a\hat{0})
\ee
is the simplest choice to obtain a non-vanishing trace in spin space.
It is the (local) N\"other charge of the invariance of the action under
the transformation 
\be
  \label{e:spin}
        \delta \heavy(x) = \omega(\vecx) \sigma_k \heavy(x)\,,\qquad
        \delta \heavyb(x) = -\omega(\vecx) \heavyb(x) \sigma_k \,, \qquad
        \sigma_{k} \equiv - \frac12 \epsilon_{ijk}\sigma_{ij}\,,
\ee
with infinitesimal space-dependent parameter $\omega(\vecx)$ and 
$[\sigma_k,\sigma_l]=i \epsilon_{klm}\sigma_m$. The Ward identities
derived from this invariance imply that $\sk(x)$ is not 
renormalized. 
Thus no additional factors are necessary in \eq{e:zspindef}. 

We have formulated the renormalization
condition in terms of correlation functions of local operators
to make it obvious that
the standard theory of renormalization 
including $\Oa$-improvement applies. However,
integrating out the static quark fields we arrive at a form which
is more natural for explicit computations, perturbative and 
non-perturbative. This step also shows immediately the connection to the
spin-dependent potentials. With the explicit form of the 
static propagator
\cite{stat:actpaper}, one obtains
\be
  {\langle\, S_1(x+\frac{L}2\hat{0}) \, \Ospin(x) \,\rangle 
      \over
      \langle\, S_1(x+\frac{L}2\hat{0}) \, S_1(x) \,\rangle  }
    = 
  {\langle\, \Tr (\pol_0(x) B_1(x)) \rangle 
      \over
      \langle\, \Tr (\pol_0(x)) \,\rangle  }\,,
  \quad
  B_1(x)=i \hat{F}_{23}(x)\,, \label{e:zspindef2}
\ee 
where the Polyakov loop operator 
\be
   \pol_\mu(x) =  W(x,\mu)W(x+a\hat{\mu},\mu)\ldots 
   W(x+(L-a)\hat{\mu},\mu) 
\ee
enters. Here $\hat F_{23}(x)$ stands for the clover leaf
discretization of the field strength tensor [20] (an alternative discretization
will also be considered in the non-perturbative computations).
Now the renormalization condition
\eq{e:zspindef} is given in terms 
of the expectation values of a (traced) Polyakov loop and 
of a (traced) Polyakov loop with the insertion
of a $B$ field.  

At this point it is useful to digress for one paragraph in
order to exhibit the relation 
to spin-dependent potentials \cite{Eichfein,Gromes}. In standard 
notation\footnote{
The spin-dependent potentials for quarks of masses $m_1,\,m_2$ and
spin operators $\vecs_1,\, \vecs_2 $ read
$$
  {\vecs_1\cdot\vecs_2 \over 3 m_1 m_2} V_4(r) + 
  {1 \over m_1 m_2} 
  \left[ {\vecx \cdot \vecs_1\, \vecx \cdot \vecs_2 \over r^2} - 
         {\vecs_1\cdot\vecs_2 \over 3}  
  \right] V_3(r)\,. 
$$
}
they can be defined as (periodic boundary conditions 
in all directions, $r^2=x_1^2+x_2^2+x_3^2$)
\bes \label{e:vspin} \nonumber
   {x_1 x_2 \over r^2}\vthreesf(r,\mu) &=&
  [\zmagsf(1/\mu)]^2  \lim_{L_0\to\infty} a\sum_{x_0}
  {\langle\, 
    \Tr (\pol_0(0) B_1(0))  \Tr (\pol_0(x)^\dagger B_2(x))\rangle 
      \over
    \langle\, \Tr\pol_0(0) \Tr \pol_0(x)^\dagger \,\rangle  } 
 \\ \nonumber
   {\vfoursf(r,\mu)-\vthreesf(r,\mu) \over 3} &=&
  [\zmagsf(1/\mu)]^2  \lim_{L_0\to\infty} a\sum_{x_0}
  {\langle\, 
    \Tr (\pol_0(0) B_1(0))  \Tr (\pol_0(x)^\dagger B_1(x))\rangle 
      \over
    \langle\, \Tr\pol_0(0) \Tr \pol_0(x)^\dagger \,\rangle  }    
\ees
Translating from the potential renormalized in the SF scheme
at renormalization scale $\mu$ to an 
RGI potential works just as explained in \sect{s:rgi}. 
In contrast to the static potential where a difficult
to determine additive renormalization
results from $\delta m_W$, there is no additive renormalization
in the above equations; $\delta m_W$ drops out. Note that we have nothing
to add to the phenomenological relevance of such 
potentials \cite{reviews:pnrqcd}.
Rather we remark that these objects, computed recently in \cite{pot:KK},
have to be renormalized with $\zmag$, which arises in HQET. 

Returning to our renormalization condition, we note that it is natural
to use the equivalence of all coordinates in Euclidean space to
switch to the usual SF boundary conditions,
where
\bes
  \label{e:zspindef3}
  \zmagsf(L) 
  {L^2\langle\, \Tr (\pol_3(x) E_1(x)) \rangle 
      \over
      \langle\, \Tr \pol_3(x) \,\rangle  }
  &=&
  \left.{L^2 \langle\, \Tr (\pol_3(x) E_1(x)) \rangle 
      \over
      \langle\, \Tr \pol_3(x) \,\rangle  }\right|_{g_0=0}
      \,,\text{ at }x_0={L_0\over 2}
  \\ \nonumber
  E_1(x)=i \hat{F}_{01}(x)\,, && \text{Dirichlet boundary
  conditions in time.}
\ees
From now on we retain these boundary conditions 
and this coordinate system.
The tree-level value is 
\be
  \left.{L^2 \langle\, \Tr (\pol_3(x) E_1(x)) \rangle 
      \over
      \langle\, \Tr \pol_3(x) \,\rangle  }\right|_{g_0=0}
  = {\pi \over 6} {1+\sqrt{3} \over 2 -\sqrt{3}} + \rmO((a/L)^4)\,.
\ee
Corresponding formulae for finite $a/L$, which are used in the 
non-perturbative definition in order to assure 
$\zmag=1$ at tree-level, are given in eqs.(\ref{e:PEtree}, \ref{e:P3tree}).

\subsection{The 2-loop anomalous dimension \label{s:AD}}

Our strategy for computing the RGI renormalization
is to non-perturbatively evaluate $\zmag(1/\mu)$ up to
$\mu=\rmO(100\,\GeV)$ and then evaluate 
$\Ospinrgi/\Ospin(\mu)$ in perturbation theory. 
In the latter step 
the anomalous dimension $\gamma(\gbar)$,
defined by the renormalization group equation
\be
  \mu{{\partial}\over{\partial\mu}}\OspinSF=\gamma(\gbarSF)\OspinSF\,,
\ee
is needed including the 2-loop term, in order to 
have a negligible perturbative uncertainty. 
We calculated the coefficient
$\gamSF_1$ in the expansion
\be
  \gamSF(\gbarSF)=-\gbarSF^2(\gamma_0+\gamSF_1\gbarSF^2+\ldots)
\ee
by conversion from the $\msbar$ scheme to the
SF scheme. 
The 2-loop anomalous dimension in the $\msbar$ scheme,
\be
  \gamMSbar_1=(\frac{17}{2} - \frac{13}{12}\nf)/(32\pi^4)\,,
\ee
is known
from~\cite{stat:eichhill_1m,Falk:1991pz,HQET:sigmabI,HQET:sigmabII}.

For the relation between the two schemes,
we use the connection of the operators in the lattice minimal 
subtraction scheme (``lat'') and the $\msbar$ scheme of dimensional 
regularization on the one hand and 
performed a 1-loop computation of the 
SF renormalization factor in the lattice regularization on the other hand.
The latter, new, computation is detailed in \app{s:pert}. Here we just quote
the results and combine them to obtain $\gamSF_1$. 

At 1-loop order, the operator in the lattice minimal 
subtraction scheme is given by
\bes
  \Ospinlat(\mu) = [1-g_0^2 \gamma_0 \ln(\mu a)] \Ospin
\ees
with the bare operator $\Ospin$ defined earlier. It is related
to the operators in the SF scheme and the $\msbar$ scheme by
finite renormalizations,
\bes
  \OspinSF(\mu) = \chi_{\rm SF,lat}(\glat^2(\mu))\Ospinlat(\mu) \,,
  \quad 
  \OspinMS(\mu) = \chi_{\rm \msbar,lat}(\glat^2(\mu))\Ospinlat(\mu) \,.
\ees
For our purposes it suffices to know the expansions up to order $g^2$,
\bes
  \chi_{a,b}(g^2) = 1 + \chi_{a,b}^{(1)}\,g^2+\ldots\,,
\ees
where 
\bes
  \chi_{\rm \msbar,lat}^{(1)} = 0.3824
\ees
has been computed\footnote{Taking into account the 
  discussion in \cite{stat:eichhill_za}, the value
  of $e$ should be reduced from $e=24.48$ in \cite{stat:ospin_flynn}
  to $e=4.53$. In the notation of \cite{stat:ospin_flynn}
  we have $\chi_{\rm \msbar,lat}^{(1)} = 
  {1\over 16 \pi^2}[C_f (D_c-e+4\pi^2)
  +C_A(D_a+D_b-D_c/2)]$. We thank Jonathan Flynn for 
  clarifying this point. When numerical uncertainties are not written
  explicitly,  they are estimated to 
  be at most of order 2 on the last digit.
  } 
by Flynn and Hill \cite{stat:ospin_flynn} and 
\bes\label{eq:Chi_SF_lat_res}
  \chi^{(1)}_{\rm SF,lat}=0.3187016(1)-0.027448(1)\,\nf
\ees
is obtained in \app{s:pert}. These are combined to
\bes
  \chi_{\rm SF,\msbar}^{(1)} = 
  \chi_{\rm SF,lat}^{(1)}-\chi_{\rm \msbar,lat}^{(1)}
  =-0.0637 - 0.0275\,\nf \,.
\ees
Note that the $\nf$-dependent part depends on the 
value $\theta$ chosen for the spatial boundary conditions
of the quark fields in \eq{e:zspindef3}. The above result refers to $\theta=-\pi/3$,
for which the expectation values in \eq{e:zspindef3} can be
shown to be real \cite{SF:symmetries}. 

The last missing ingredient is the relation between the couplings,
(at the same renormalization scale)
\be
  \gbarSF^2=\chi_{\rm g}\gbarMSbar^2, \qquad
  \chi_{\rm g}=1+\chi_{\rm g}^{(1)}\gbarMSbar^2+O(\gbarMSbar^4)\,,
\ee
with the 1-loop coefficient~\cite{alpha:SU3,pert:1loop}
\be
  \chi_{\rm g}^{(1)}=-{1\over 4\pi}(c_{1,0}+c_{1,1}\nf)
  \,,\quad  c_{1,0}=1.25563(4)\,,\; c_{1,1}=0.039863
\ee
for the SF-coupling defined at $\theta=\pi/5$ \cite{pert:1loop}. 

Analogous to~\cite{mbar:pert}, where the discussion is carried through for
the anomalous dimension of quark masses, we then obtain the desired result
\be
  \gamSF_1=\gamMSbar_1+2b_0\,\chi_{\rm SF,\msbar}^{(1)}-
  \gamma_0\chi_{\rm g}^{(1)} = -0.00236 - 0.00352\,\nf + 0.00023\,\nf^2\,.
\ee
Our computation used the lattice regularization
but the result is regularization independent.

\section{Step scaling functions \label{s:SF}}
\subsection{Definition}
The anomalous dimension, which we just obtained in 2-loop 
approximation, describes the change of the operator $\OspinSF$
under an infinitesimal change in the renormalization scale.
For numerical, non-perturbative computations one
considers finite changes of the scale, typically by
a factor of two.  These define the step scaling function
$\sigmamag(u)$ via
\be
  \OspinSF(\mu) = \sigmamag(\gbar^2(1/\mu))\, \OspinSF(2\mu)\,.
\ee
It is given by the continuum limit
\be
  \sigmamag(u) = \lim_{a/L\to0} \Sigmamag(u,a/L) \label{e:contlim}
\ee
of the lattice step scaling function
\be\label{eq:Sigma_spin}
   \Sigmamag(u,a/L) = \left.
     {\zmagsf(2L)  \over \zmagsf(L) }\right|_{\gbar^2(L)=u\,,\;m=0}\,.
\ee 
Here $\gbar^2(L)$ denotes the SF coupling  as before and
the condition $m=0$ means that the renormalization condition 
is imposed at vanishing quark mass. In our numerical implementation
the latter does not play a r\^ole since we  work in the 
pure gauge theory. It is important to understand whether
a given renormalization condition leads to small or large
$a$-effects. This determines whether 
the limit in \eq{e:contlim} can be taken by an extrapolation
from the accessible lattices.
A first understanding can be sought in perturbation theory. 

\subsection{Lattice artefacts in perturbation theory}

We define the relative lattice artefacts as
\be
 \delta(u,a/L) =  {\Sigmamag(u,a/L) - \sigmamag(u) \over \sigmamag(u)} = 
 \delta_1(a/L)\, u + \rmO(u^2)\,. 
\ee
A term $\delta_0(a/L)$ is absent since we defined
$\zmagsf(L)$ such that it is one at tree-level for
any value of $a/L$. The 1-loop term may be expanded
in $\nf$,
\be
 \delta_1(a/L) = \delta_{1,0}(a/L) + \nf\, \delta_{1,1}(a/L) \,.
\ee
With all improvement terms, including the boundary improvement 
term $\ct$ \cite{alpha:SU3}, set to their proper
perturbative values, the 1-loop cutoff effects 
turn out to be rather small. We show them in \tab{t:delta} 
for the Eichten-Hill action for the static quarks ($W(x,0)=U(x,0)$),
the plaquette gauge action and the $\Oa$-improved fermion action \cite{impr:pap1}.

\TABLEH{
\begin{tabular}{crr}
\hline \\[-1.75ex]
$L / a$   &   $\delta_{1,0}(a/L)$  & $\delta_{1,1}(a/L)$ \\[1ex]
\hline \\[-1.75ex]
 6  &-0.000236 & 0.013742  \\
 8  &-0.000165 & 0.005791 \\
 10 &-0.000106 & 0.003026 \\
 12 &-0.000072 & 0.001876\\
 14 &-0.000051 & 0.001296\\
 16 &-0.000038 & 0.000956\\
    &          &         \\[-1.75ex]
\hline
\end{tabular}
\caption{\footnotesize 
Lattice spacing effects of $\Sigmamag$ in 1-loop perturbation theory, see text.
}\label{t:delta}
}

\subsection{Non-perturbative results for $\nf=0$}

We carried out pure gauge theory simulations to determine
$\Sigmamag$ for different couplings $u$, resolutions 
$1/12 \leq a/L \leq 1/6$ and also for different 
discretizations of the HQET action and the operator
$F_{\mu\nu}$. Tables of the numerical results are found
in \app{s:MC}. 

\FIGURE{
\includegraphics*[width=7.5cm]{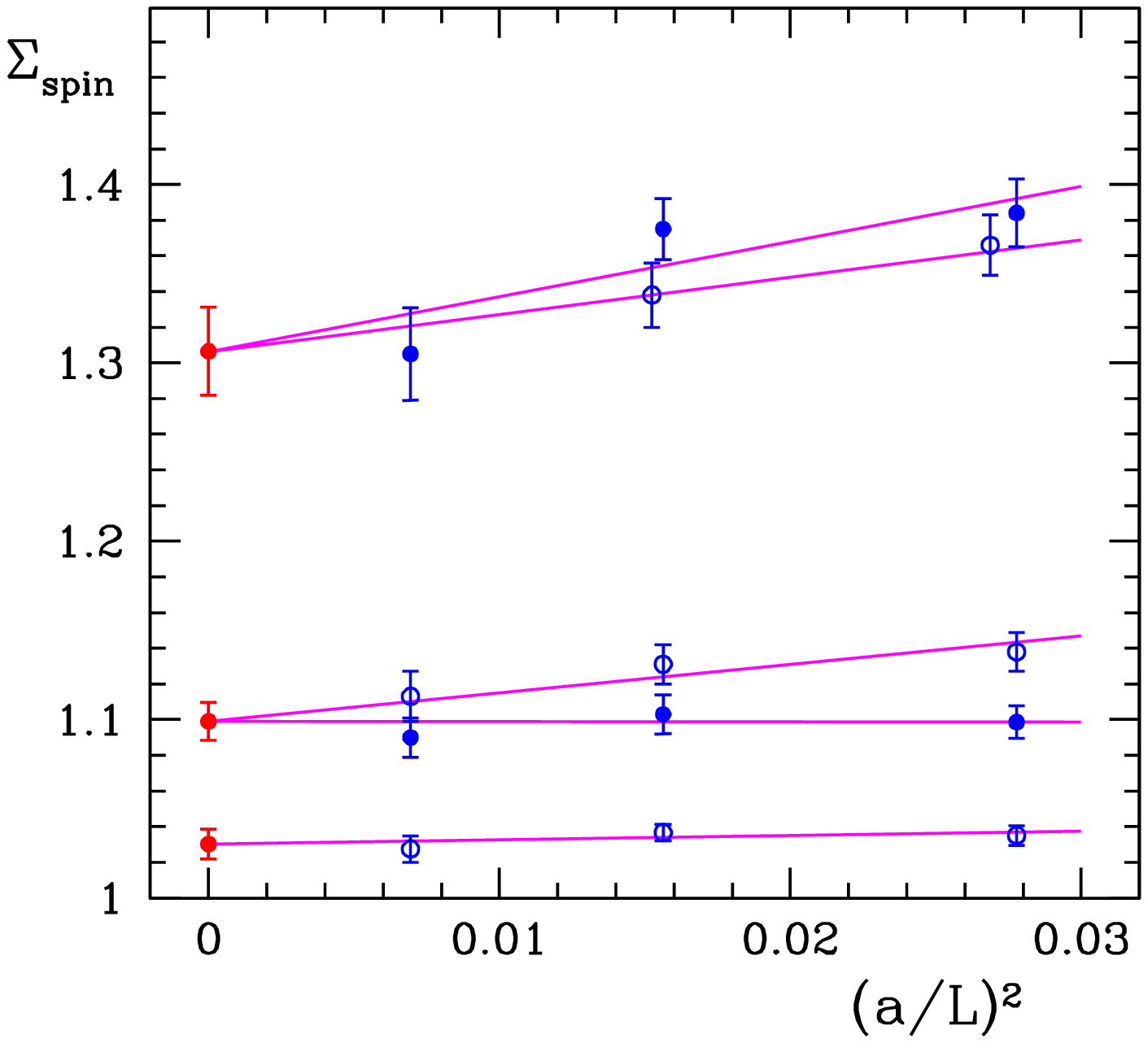}
\caption{\footnotesize
Examples of continuum extrapolations of $\Sigmamag$ 
for $u=$1.243, 2.77 and 3.48.
Filled symbols indicate that $F_{\mu\nu}$ was defined
as $\hat{F}_{\mu\nu}$ but with the link variables replaced
by HYP2 links. The data at the largest coupling has 
$\ct$ at 2-loop precision, otherwise it is set to the
1-loop value.
}\label{f:Sigma}
}

In \fig{f:Sigma} we show the $a$-dependence
for a few couplings and choices of discretizations. 
Lattice artefacts are moderate in general 
and a simple
continuum extrapolation with an ansatz 
$\Sigma = \sigma + (a/L)^2\,\rho$,
separately for each value of $u$ seems 
justified.\footnote{We also checked other extrapolations,
see e.g. \cite{mbar:pap1,mbar:nf2}. They give compatible 
results.}
In cases where more than one discretization was
simulated at the same value of $u$, a constrained
extrapolation (common $\sigma$ but discretization-dependent 
$\rho$) was performed. This simple analysis yields
the continuum step scaling functions in \tab{t:sigma}. 

\TABLEH{
\begin{tabular}{clrccc}
\hline \\[-1.75ex]
$u$  & $\sigmamag(u)$  &  $\rho(u)$  & action & $\hat{F}_{\mu\nu}$ & $n$-loop $\ct$ \\[1ex]
\hline \\[-1.75ex]
0.8873&1.0235(68)&0.02(36) & &    & \\[0.3ex]
0.9944&1.025(11) &-0.15(53)& EH   &   & \\[0.3ex]
1.2430&1.0302(82)&0.24(42) & &    & \\[0.3ex]
1.3293&1.043(12) &-0.66(55)& EH   &   & \\[0.3ex]
1.5553&1.0418(89)&0.27(43) & &    & \\[0.3ex]
1.8811&1.075(11) &-0.44(52)& &    & \\[0.3ex]
2.1000&1.052(14) &1.15(69) & &    & \\[0.3ex]
2.4484&1.089(14) &-0.31(68)& &    & \\[0.3ex]
      &          & 1.60(60)& &    & \\[-1.3ex]
2.770 &1.099(11) &         & &    &  \\[-1.3ex]
      &          &-0.01(55)& &HYP2&  \\[0.3ex]
3.480 &1.300(34) &5.2(1.6) & &HYP2& \\[0.3ex]
3.480 &1.308(29) &3.0(1.4) & &HYP2&2 \\[0.3ex]
      &          &2.1(1.2) & &    &2\\[-1.3ex]
3.480 &1.306(25) &         & &      &\\[-1.3ex]
      &          &3.1(1.3) & &HYP2&2\\
      &          &         & &     & \\[-1.75ex]
\hline
\end{tabular}
\caption{\footnotesize 
Continuum step scaling function $\sigmamag$ and
coefficients of $(a/L)^2$ in the continuum extrapolations. 
The standard discretization is the 
HYP2 action of \cite{stat:actpaper}, the clover leaf 
operator $\hat{F}_{\mu\nu}$ and the 1-loop value of $\ct$. Deviations
from this rule are indicated: EH refers to the Eichten-Hill
static action and the label HYP2 in the column $\hat{F}_{\mu\nu}$ 
means that the links appearing in $\hat{F}_{\mu\nu}$ are replaced
by HYP2 links.
}\label{t:sigma}
}

A good agreement with perturbation theory is seen at
weak couplings, say $\gbar^2 < 2$, where, however, the
non-perturbative results are not accurate enough to
distinguish between the 2-loop anomalous dimension
and the 1-loop one. On the other hand, for couplings
$\gbar^2 \approx 3$ perturbation theory breaks down entirely 
and the inclusion of the 2-loop anomalous dimension 
brings the perturbative curves even further away 
from the MC results.

\FIGURE{
\includegraphics*[width=8.2cm]{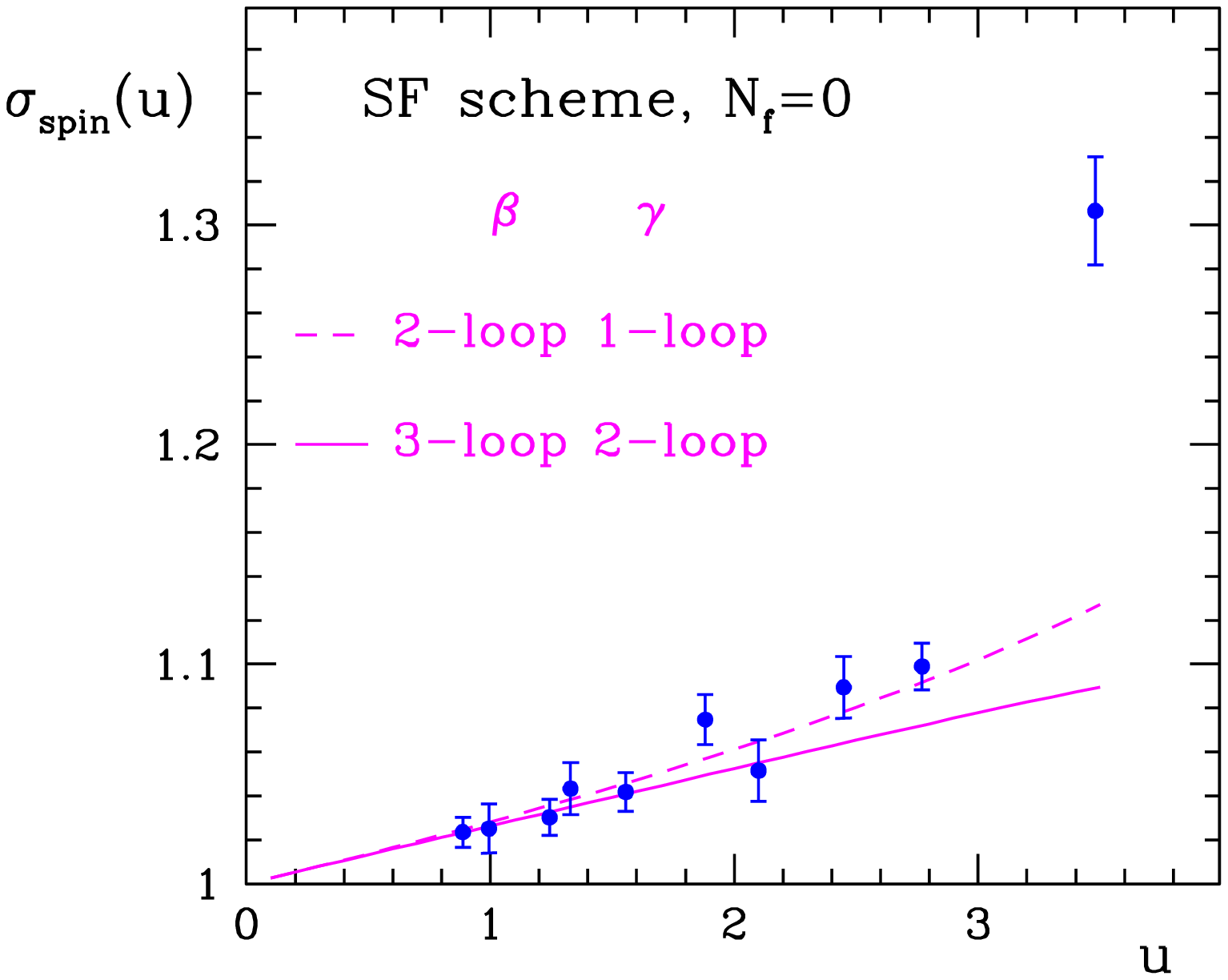}
\caption{\footnotesize
The step scaling function compared to the perturbative
prediction obtained by integration of the perturbative
RG equations, truncating the $\beta$ function at
3(2)-loop order and the anomalous dimension at 2(1)-loop order.
}\label{f:sigma}
}

\section{Non-perturbative scale dependence for $\nf=0$ \label{s:running}}

Through an iterative application of the (inverse) step scaling 
function we now determine the non-perturbative 
scale dependence of the operator in the SF scheme.
We first represent the numerical data
of \tab{t:sigma} as well as the data for the 
step scaling function of the coupling, $\sigma(u)$
(Table~A.2 of \cite{mbar:pap1} and Table~4 of \cite{alpha:SU3}) by 
smooth interpolations 
\bes
  \label{e:fitsig}
  \sigma(u) &=& u + \sum_{i=0}^4s_i\,u^{i+2}  \,,
  \\
  \label{e:fitsigm}
  \sigmamag(u) &=& 1 + m_0\,u + m_1\,u^2 + m_2\,u^3\,   
\ees
for $u\leq 2.8$. Here $s_0=2 \ln(2)\,b_0\,,\; s_1=s_0^2+ 2 \ln(2)\,b_1\,,\;m_0=\ln(2)\,\gamma_0$
are fixed by perturbation theory, while the other coefficients
are free fit parameters. With these parameterizations we then 
solve the recursions \footnote{
$\Phi$ stands for any matrix element of the operator $\Ospin$, 
for example the matrix element $\lambda_2$. 
}
\bes
  u_0=\gbar^2(\lmax) = 3.48\,,\quad \sigma(u_{k+1})=u_k\,,&\Rightarrow&
  \gbar^2(2^{-k}\lmax)=u_k\,, \nonumber \\[-3ex] \\[-1ex]\nonumber
  w_0=1\,,\quad w_{k+1}=w_k/\sigmamag(u_{k+1}) &\Rightarrow& 
  {\Phisf(2^k/\lmax) \over \Phisf(1/\lmax)} =w_k\,. 
\ees
Errors are propagated through the parameterization and recursion 
and it is checked
that changing the number of fit parameters in \eq{e:fitsig} and \eq{e:fitsigm}
does not alter the results significantly. Next we apply 
\be \label{e:int}
   {\Phirgi \over \Phisf(\mu)} = \left[\,2b_0 \gbar^2(\mu)\,\right]^{-\gamma_0/2b_0}
                   \exp\left\{-\int_0^{\gbar(\mu)} \rmd g
                     \left[\,{ \gamma^{\rm SF}(g) \over\beta^{\rm SF}(g)}
                           -{\gamma_0 \over b_0 g}\,\right]
                     \right\} \,
\ee
for $\mu=2^k/\lmax$ and the perturbative approximations to the
$\beta$-function and anomalous dimensions 
and obtain $\Phisf(1/\lmax)/\Phirgi$. Finally we form 
$\Phisf(1/(2\lmax))/\Phirgi = \sigmamag(u_0)\, \Phisf(1/\lmax)/\Phirgi$ 
with $\sigmamag(u_0)$ from \tab{t:sigma}.
For not too small $k$, the final result is independent of it; 
the use of perturbation theory is safe in that region. 
Taking $k=6$, the 2-loop approximation
for the anomalous dimension and the 3-loop approximation for the 
$\beta$-function we arrive at
\bes
  \label{e:univ1} \Phisf(\mu)/\Phirgi &=& 0.759(17)\,, \quad \text{at } \mu=1/\lmax \,, \\
  \label{e:univ2} \Phisf(\mu)/\Phirgi &=& 0.992(29)\,, \quad \text{at } \mu=1/(2\lmax) \,.
\ees
\Fig{f:running} compares the non-perturbative running of 
$\Ospin$ to perturbation theory. 

\FIGURE{
\includegraphics*[width=9.5cm]{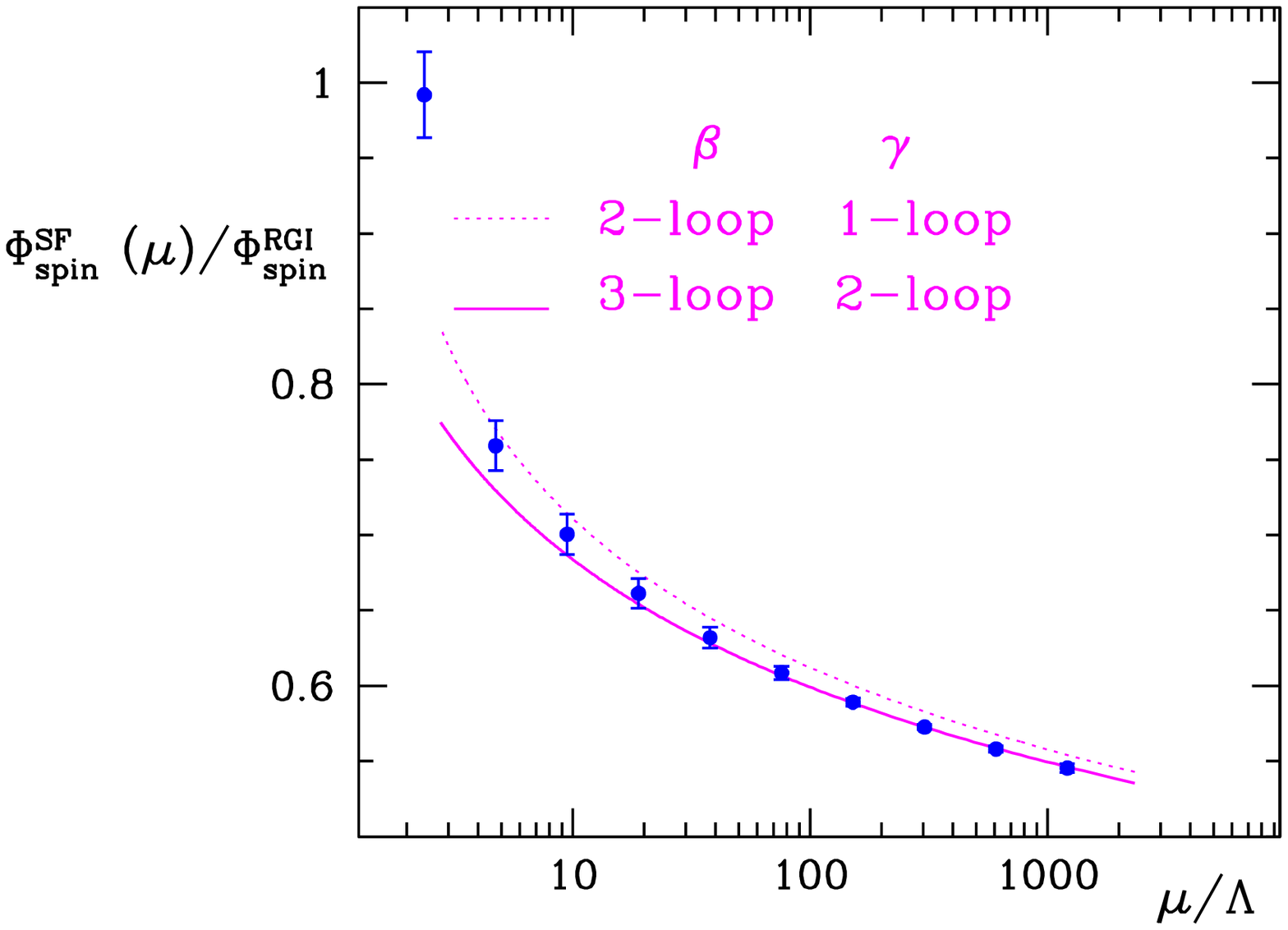}
\caption{\footnotesize
Scale dependence of $\Ospin$ in the SF scheme. 
}\label{f:running}
}

\subsection{The relation of bare and renormalization group invariant 
operator \label{s:zfact}}

The universal result \eq{e:univ2} has to be combined with values 
of $\zmagsf(L)$ 
at $L=2\lmax=1.436\,r_0$ \cite{pot:r0_SU3} which depend on the bare
coupling and lattice action, to form 
\be
  \zmagrgi = \zmagsf(L) \times {\Phirgi \over \Phisf(1/L)}
\ee
for the respective action. The numerical values of
\tab{t:zlref} are well represented by
\bes
\zmagsf(2\lmax) &=& 2.58 + 0.14\,(\beta-6)-0.27\,(\beta-6)^2 \,\quad\text{EH action}\,, \\
\zmagsf(2\lmax) &=& 2.59 + 0.11\,(\beta-6)-0.34\,(\beta-6)^2 \,\quad\text{HYP2 action}\,.
\ees
These interpolations may be used in the interval $6.0\leq\beta\leq6.5$ with an error
of about 1\%. While this error ought to be taken into account before 
the continuum extrapolation of the renormalized matrix elements, 
the uncertainty in \eq{e:univ2} applies additionally in the continuum limit.

\TABLEH{
\begin{tabular}{cccr}
\hline \\[-1.75ex]
$\beta$ & $L / a$   & action  & $\zmagsf$ \\[1ex]
 & & &  \\[-1.75ex]
\hline\\[-1.75ex]
6.0219& 8   &  EH &2.585(19)   \\ 
6.1628& 10  &  EH &2.602(24)   \\
6.2885& 12  &  EH &2.602(24)   \\
6.3992& 14  &  EH &2.589(29)   \\
6.4956& 16  &  EH &2.593(35)   \\
 & & &  \\[-1.75ex]
\hline \\[-1.75ex]
6.0219& 8   & HYP2 &2.593(21)  \\
6.1628& 10  & HYP2 &2.589(26)  \\
6.2885& 12  & HYP2 &2.589(26)  \\
6.3992& 14  & HYP2 &2.585(21)  \\
6.4956& 16  & HYP2 &2.553(22)  \\
 & & &  \\[-1.75ex]
\hline
\end{tabular}
\caption{\footnotesize 
Renormalization factor at the matching scale. 
In all cases $\hat{F}_{\mu\nu}$ is the standard
clover operator.}\label{t:zlref}
}

\section{First applications \label{s:appl}}

We illustrate the usefulness of our result with two
sample applications.

\subsection{Spin splitting}

First we
take numbers for the bare $\lambda_2$ which have been 
reported in the literature. Unfortunately they exist only
for $\beta=6.0$, which corresponds to $a\approx 0.1\,\fm$.
The more recent evaluations are (the light quark has the
mass of the strange quark)
\bes
   \text{Ref.~\cite{hqet:spinsplitt_gms}:}  && a^2 \lambda_2^{\rm bare} = 0.0100(19)\,,\\
   \text{Ref.~\cite{hqet:spinsplitt_jlqcd}:} && a^2 \lambda_2^{\rm bare} = 0.0138(15)\,.
\ees
The authors of \cite{hqet:spinsplitt_gms,hqet:spinsplitt_jlqcd} then estimate the mass
splitting as
\bes
   \text{Ref.~\cite{hqet:spinsplitt_gms}:}  && \Delta m^2 =0.28(6)(?)\, \GeV^2\,,\\
   \text{Ref.~\cite{hqet:spinsplitt_jlqcd}:} && \Delta m^2= 0.36(4)(?)\,\GeV^2\,,
\ees
where the renormalization factor is taken from perturbation
theory using a boosted coupling \cite{stat:bbstar}
and by tadpole improved perturbation theory 
\cite{hqet:spinsplitt_jlqcd}\footnote{Also 
somewhat different values for the lattice spacing were used by the 
two groups.}.
 With 
$\Cspin(\Mbeauty/\Lambda_\msbar)=1.15$, $\Mbeauty=6.76(9)\,\GeV$
\cite{hqet:pap4,lat06:damiano} 
and
$\zmagrgi=2.6\,, a=1/(2\,\GeV)$ we find from \eq{e:splitt2} 
\bes
   \text{Ref.~\cite{hqet:spinsplitt_gms} and NP $\zmagrgi$:}  && 
   \Delta m^2 = 0.38(7)(?)\,\GeV^2\,,\\
   \text{Ref.~\cite{hqet:spinsplitt_jlqcd} and NP $\zmagrgi$:} && 
   \Delta m^2=0.53(6)(?)\GeV^2\,.
\ees
In all these estimates the additional uncertainty marked as (?) 
refers to lattice artefacts,
namely the fact that a continuum limit has not been taken, and of course 
to the missing dynamical quark determinant. The experimental mass 
splitting is $\Delta m^2= 0.497\,\GeV^2$.

\subsection{Renormalization factor for spin-dependent potentials}

In phenomenological applications of the spin-dependent potentials, the 
standard renormalization scheme is $\msbar$, as in
\cite{stat:eichhill_1m,Falk:1991pz,HQET:sigmabI,HQET:sigmabII}.
We apply \eq{e:int} in the $\msbar$ scheme with 
$\nf=0$, $\Lambda_{\msbar}^{(0)}=238\,\MeV$ \cite{mbar:pap1},
the 4-loop $\beta$-function and the 2-loop
anomalous dimension. This yields
\bes
  \zmag^{\msbar}(2\,\GeV) &=& 0.756(18) \times \zmagrgi(g_0) \,, \\
  \zmag^{\msbar}(4\,\GeV) &=& 0.706(13) \times \zmagrgi(g_0)  \,,
\ees
where the cited error bar is half of the change when one uses
the 1-loop anomalous dimension instead. Remember from
 \sect{s:scheme} that the square of
this renormalization factor enters the potentials.
As an illustration, for the standard Eichten-Hill action 
for the
static quarks, at $g_0=1$ ($\beta=6.0$), 
we then have $Z_{\msbar}(2\,\GeV) \approx 0.756 \times
2.58 \times 0.99 = 1.93$ or a somewhat smaller
number at larger $\mu$.  One can compare this at $g_0=1$ to the 
renormalization factors used in \cite{pot:KK}.
A tree-level tadpole-improved factor is $Z_{\rm tad}=1.684$
and the 
Huntley-Michael factor\footnote{The prescription of
\cite{pot:HMfact} yields a small $r$-dependence, which
is not present in the standard renormalization
of local operators.} is around 
$Z_{\rm HM}=1.62$. Since a renormalization scheme and scale are 
not specified in
these procedures, there is no reason to
expect a closer agreement. 

\section{Conclusions \label{s:concl}}

We have presented yet another example that 
the non-perturbative renormalization programme
using recursive finite size techniques \cite{alpha:sigma,mbar:pap1}
can be carried out also in difficult cases. 
Four-fermion operators were renormalized
successfully in \cite{4ferm:nf0,4ferm:pert}, 
and those containing static quarks in
\cite{stat:zbb_pert,lat06:carlos}. Here we have treated
the case of an operator with static quarks and 
gluon fields. 

In several cases quite significant deviations
from perturbation theory had already 
been observed at intermediate to low renormalization
scales \cite{zastat:pap3,zastat:nf2,4ferm:nf0},
but the present case is the strongest example in that respect 
(\fig{f:sigma}, \fig{f:running}). 

Our example in the previous section illustrates 
that the 
non-perturbative $\zmagrgi$ has a rather big effect. 
Although the quenched estimates for $\Delta m^2 = \mbsstar^2-\mbs^2$
are in rough agreement with the experimental mass splitting 
$\Delta m^2= 0.497\,\GeV^2$
after the non-perturbative $\zmagrgi$ is used, it now remains
to improve the precision of the bare matrix element
as well as to obtain it at smaller lattice spacings
in order to see whether the quenched approximation does indeed
give a reasonable estimate of the spin splitting. 
We also emphasize that there is a remaining uncertainty
in the use of (continuum) perturbation theory for 
$\Cspin(\Mbeauty/\Lambda_\msbar)$. This can be significantly 
reduced by a computation of the associated 3-loop anomalous 
dimension, but also an entirely non-perturbative matching
of HQET and QCD is promising \cite{hqet:pap1,hqet:pap4}.

\vspace{0.4cm}
{\bf Acknowledgements.} 
We acknowledge useful discussions with N. Brambilla, 
M. Della Morte, J. Flynn, P. Marquard, J. Pilcum, J.~Soto, 
M. Steinhauser, A. Vairo, S.~Takeda and U.~Wolff.
We thank NIC  for allocating computer time on the APEmille
computers at DESY Zeuthen to this project and the APE group for its help. This
work is supported by the  Deutsche Forschungsgemeinschaft 
in the SFB/TR~09 , by 
the European community through 
EU Contract No.~MRTN-CT-2006-035482, ``FLAVIAnet''
and by funds provided by the U.S. Department of Energy under 
cooperative research agreement DE-FC02-94ER40818.

\begin{appendix}
\hyphenation{ per-tur-bative renor-ma-li-zation }

\newcommand{\beqa}{\begin{eqnarray}}
\newcommand{\eeqa}{\end{eqnarray}}
\newcommand{\beq}{\begin{equation}}
\newcommand{\eeq}{\end{equation}}
\newcommand{\re}{\mathop{\rm Re}}
\newcommand{\Hc}{{\rm H.c.\ }}
\newcommand{\sss}{{s}}
\newcommand{\nn}{\nonumber \\}
\newcommand{\dg}{^\dagger}
\newcommand{\xp}{x_\perp}
\newcommand{\Lp}{L_\perp}
\newcommand{\sig}{\sigma}
\newcommand{\tsig}{\tilde\sigma}
\newcommand{\<}{\langle}

\renewcommand{\>}{\rangle}
\renewcommand{\eq}[1]{eq.~(\ref{#1})}
\renewcommand{\Eq}[1]{Eq.~(\ref{#1})}
\renewcommand{\eqs}[1]{eqs.~(\ref{#1})}
\renewcommand{\Eqs}[1]{Eqs.~(\ref{#1})}
\renewcommand{\fig}[1]{Fig.~\ref{#1}}
\renewcommand{\Fig}[1]{Figure~\ref{#1}}
\renewcommand{\sect}[1]{Sect.~\ref{#1}}
\renewcommand{\Sect}[1]{Section~\ref{#1}}
\renewcommand{\tab}[1]{Tab.~\ref{#1}}

\newcommand{\la}{\label}
\newcommand{\figs}{Figs.~}
\newcommand{\n}{$N$}
\newcommand{\im}{\mathop{\rm Im}}
\newcommand\mycaption[1]{\caption{\footnotesize \sf #1}}
\newcommand{\toprule}{\hline}
\newcommand{\midrule}{\hline}
\newcommand{\bottomrule}{\hline}

\newcommand{\gounder}{g_0^{\phantom{\dagger}}}

\section{One-loop computation\la{s:pert}}

Our aim is to compute the expectation value of a general Wilson loop at one-loop
order in the Schr\"odinger functional (SF), bearing in mind that we are finally interested
in the computation of the expectation value of a clover operator $\hat{F}_{\mu\nu}$ 
inserted into a Polyakov loop, enabling us to obtain 
the two-loop anomalous dimension of the chromo-magnetic operator 
in the SF scheme from its value in the lat scheme.
Due to the space-time locality of such an observable,  it will be advantageous
to compute the gluon loops in $x$ space, while the contribution of tadpoles
is proportional to the zero-momentum gluon propagator. Quite a lot of notation will be 
introduced, but the formulae presented here are suitable for the automated computation
of arbitrary Wilson loops.

The lattice spacing is set to one in this appendix. Space-time 
indices run from 1 to 4, the latter being associated with time.
For everything else we reuse the notation of \cite{thesis:skurth}, referred
to as (K) in what follows, except that a twiddle
on the color components of gluon, ghost and quark fields is dropped.
The reader is assumed to be familiar with chapters 3, 4 and 5 of (K).
Up to one-loop order, the observable ${\cal O}$ has the perturbative expansion
(K:4.39)
\be
\< {\cal O}\> = {\cal O}^{(0)}  + g_0^2 \left[ \< {\cal O}^{(2)} \>_0 - 
                      \< {\cal O}^{(1)} S^{(1)} \>_0  \right]  + {\rm O}(g_0^4),
\ee
where $\<\dots\>_0$ is the expectation value with respect to the free
part $S^{(0)}$ of the action and ${\cal O}^{(k)}$ is defined by
${\cal O} = {\cal O}^{(0)} + {\cal O}^{(1)}g_0 + {\cal O}^{(2)}g_0^2 +\dots $.

The (constant, Abelian) background field induced by the non-trivial boundary 
conditions takes the value $V_\mu(x)$.
We use the basis of the $su(3)$ Lie algebra
\be
q_\mu(x) = \sum_{a=1}^8  q^a_\mu(x) I^a
\ee
defined in (K:App.~A) as well as the Fourier representation 
\beqa
 q_4^a(x) & =& \frac{1}{L^3} \sum_{\bf p} e^{i {\bf p}\cdot {\bf x} } ~
     q_4^a({\bf p}, x_4). \\
 q_k^a(x) & =& \frac{1}{L^3} \sum_{\bf p} e^{i {\bf p}\cdot {\bf x} } ~ 
    e^{i(p_k+\phi_a(x_4))/2}  q_k^a({\bf p}, x_4).  
\eeqa
The gluon propagator in mixed representation used by (K) has the form
\be
\<  q_\mu^a({\bf p},x_4) ~  q_\nu^b({\bf p}',y_4) \>_0 
= \delta_{b\bar a} L^3 \delta_{{\bf p} + {\bf p'}} ~ D_{\mu\nu}^a({\bf p};x_4,y_4).
\ee
Kronecker symbols such as $\delta_{\bf p}$ or $\delta_{\mu-4} $ 
carrying a single index are shorthands for
 $\delta_{\bf p, 0}$ and  $\delta_{\mu,4}$ respectively. 
%
\subsection{Parameterization of the observable\la{sec:general}}
%
In order to compute the expectation value of an arbitrary Wilson loop at one-loop order,
we parameterize the loop by a starting point $x^{({\rm start})}$ and an ordered list $\vec\ell$ of 
length $\ell$. The entries of the list are directions $\mu_i^{\vec\ell}$, $i=1,\dots,\ell$.
These directions take non-zero integer values between $-4$ and $+4$.  An electric plaquette 
in the $(03)$ plane is thus parameterized by $\vec\ell=(3~~4~~-3~~-4)$. Clearly the loop is closed
if and only if each integer appears as many times with the $+$ sign as it does with the $-$ sign
(modulo $L$ for the space directions).
We normally drop the $\vec\ell$ in $\mu_i^{\vec\ell}$ since we will be dealing only with one path
at a time.

The sequence of points the loop goes through is obtained as follows, 
\be
x^{(1)}=x^{({\rm start})} \qquad  x^{(i+1)}=x^{(i)} ~+~  \hat{\mu}_i,\quad i=1,\dots,\ell-1.
\ee
$\hat\mu= {\rm sign}(\mu) \widehat{|\mu|}$ 
are unit vectors pointing in the four $\pm$ directions of the lattice.
We identify $\hat{4}=\hat{0}$.
At tree-level, the expectation value of the Wilson loop is
\be
W_{\vec\ell}[V] = \prod_{i=1}^\ell V(x^{(i)},\mu_i).
\ee
In general, for any 4-vector we introduce negative-index components 
\be
p_{-\mu} = - p_{\mu}.
\ee
Because of the way the path is parameterized, 
for any link variable we introduce negative-index components by imposing
\be
U(x,\mu) = U^\dagger(x+\hat\mu,-\mu),\qquad 
q_{\mu}^a(x) =  - q_{-\mu}^{a}(x+\hat\mu).
\ee
The Fourier representation is now defined for all $\mu$ as follows,
\be
q_\mu^a(x) = \frac{1}{L^3} \sum_{\bf p} 
e^{i{\bf p}\cdot {\bf x}} ~ e^{i\theta_a({\bf p},x_4,\mu)} ~
                           q_\mu^a({\bf p},x_4)
\ee
where
\be
q_{\mu}^a({\bf p},x_4)  = q_{|\mu|}^{a}({\bf p},x_4-\delta_{\mu+4} )
\ee
and 
\beqa
e^{i\theta_a({\bf p},x_4,\mu)} &=&  \left\{ \begin{array}{l@{\qquad}l} 
1 & {\rm if} ~ \mu =4 \\
e^{i(p_k+\phi_a(x_4))/2} & {\rm if} ~ \mu =k \\
-e^{i(-p_k+\phi_a(x_4))/2} & {\rm if} ~ \mu =-k \\
-1                          & {\rm if} ~ \mu =-4  \end{array}\right.  \nn
&=& {\rm sign}(\mu)\left(\delta_{|\mu|-4} + (1-\delta_{|\mu|-4}) 
     e^{i( p_{\mu}+\phi_a(x_4))/2}\right).
\eeqa

With these notations we have
\be
\< q_\mu^a({\bf p},x_4) q_\nu^b({\bf p'},y_4) \>
= \delta_{b \bar a} L^3 \delta_{{\bf p} + {\bf p'}} 
D_{|\mu||\nu|}^{a}\left({\bf p}; x_4-\delta_{\mu+4};y_4-\delta_{\nu+4}\right).
\ee

%
\subsection{Tadpoles: $  \< {\cal O}^{(1)} S^{(1)} \>_0  $}
%
The terms considered in this subsection owe their existence to the non-vanishing 
background field. Since the latter is diagonal, 
the $V$ matrices all commute two-by-two and we have
\be
\tr\{W_{\vec \ell}^{(1)} \} = \sum_{j=1}^\ell \tr\{  q_{\mu_j} (x^{(j)}) ~ W_{\vec\ell}[V] \}.
\ee
The three contributions to $S^{(1)}$ (coming from the gauge, ghost and quark terms)
are given by eqs.~(K:5.64,5.72) 
and for the quarks eqs.~(K:5.76,5.82).
After a short calculation one finds
\be
\< \tr\{W_{\vec\ell}^{(1)}\}  S^{(1)} \>_0 =
- \sum_{a\in\{3,8\}} \sum_{\mu=1}^3 \sum_{u_4} \alpha_{\vec\ell,\mu}^a(u_4) ~ T_\mu^a(u_4)\,,
\la{eq:11}
\ee
with 
\beqa
\alpha_{\vec\ell,\mu}^a(u_4) &=& 
\tr\{ I^{\bar a} W_{\vec\ell}[V] \}  \sum_{j=1}^\ell {\rm sign}(\mu_j)
 \left( \delta_{|\mu_j|-4} + (1-\delta_{|\mu_j|-4}) e^{-i\phi_a(x_4^{(j)})/2}  \right) \nn
&& D_{\mu|\mu_j|}^a\left({\bf 0};u_4,x_4^{(j)} - \delta_{\mu_j+4} \right).
\la{eq:alpha_ell}
\eeqa
and
\be
T_{\mu}^a(u_4) = T_{\mu,{\rm gluon}}^a(u_4) + T_{\mu,{\rm ghost}}^a(u_4) 
                 + \nf T_{\mu,{\rm quark}}^a(u_4).
\la{eq:tad}
\ee
$T_{\mu,{\rm gluon}}^a,~T_{\mu,{\rm ghost}}^a,~T_{\mu,{\rm quark}}^a$ are defined respectively 
by eqs.~(K:5.114), (K:5.115) and (K:5.116). Note that for the ghost and quark case, the overall 
minus sign in \eq{eq:11} arises because of the loop (ghosts and quarks are anti-commuting). 
For the gluons, the minus sign is just a 
convention chosen by (K) and is compensated by 
a minus sign in the definition of $T_{\mu,{\rm gluon}}^a$.
%
\subsection{Gluon loops: $\< {\cal O}^{(2)}\>_0$ }
%
We separately consider two contributions:
\be
{\cal O}^{(2)} = {\cal O}^{({\rm 2a})} + {\cal O}^{({\rm 2b})}.
\ee
Expanding the exponential of the gluon field to linear order, we obtain
\be
\tr\{ W_{\vec\ell}^{({\rm 2a})} \} = 
\sum_{j=1}^{\ell} \sum_{j'=j+1}^\ell 
\tr\{q^{(j)} W_{\vec\ell}(j~\sss_j| j'~\sss_{j'}) 
     q^{(j')} W_{\vec\ell}(j'~\sss_{j'}|j~\sss_j) \}
\la{eq:W2a}
\ee
where we have used the cyclicity of the trace and the shorthand
$q^{(j)} \equiv q_{\mu_j}(x^{(j)})$. 
There is also a contribution from the quadratic piece of the 
exponential of the gluon field,
\be
\tr\{ W_{\vec\ell}^{({\rm 2b})} \} = \frac{1}{2} \sum_{j=1}^\ell
\tr\{ q_{\mu_j}^2(x^{(j)}) W_{\vec\ell}[V] \} ,
\ee
which is $1/2$ of the term $j=j'$ in (\ref{eq:W2a}).

We also need the notation
$ \sss_j\equiv \frac{1}{2}(1-{\rm sign}(\mu_j)) $
and $\overline n = 1+\rm mod(n-1,\ell)$ for $n\geq1$.
Now we can formulate the definition
\be
W_{\vec\ell}(j~\sss_j | j'~\sss_{j'}) =  \left\{ \begin{array}{l @{\qquad} l}
W_{\vec\ell}[V] & {\rm if~}  
\overline{j+\sss_j} = \overline{j'+\sss_j'} {\rm ~and~} \sss_j=0\\
W(\overline{j+\sss_j} \to \overline{j'+\sss_j'}) & {\rm otherwise} 
\end{array}\right.
\ee
that invokes the parallel transporter along the loop from $x^{(j)}$ to  $x^{(j')}$:
\be
W_{\vec\ell}(j\to j') = \left\{ \begin{array}{l@{\qquad} l}
1 & {\rm if} ~ j= j' \\
\prod_{i=j}^{j'-1} V(x^{(i)},\mu_i) & {\rm if} ~ j < j' \\
\prod_{i=j}^{\ell} V(x^{(i)},\mu_i) \prod_{i=1}^{j'-1} V(x^{(i)},\mu_i)& {\rm if} ~ j > j'
\end{array}\right.
\ee
One then finds
\beqa
 \< \tr\{ W_{\vec\ell}^{({\rm 2a})} \} \>_0 =
 {1\over L^3} \sum_{j=1}^\ell \sum_{j'=j+1}^\ell \sum_{a=1}^8 \sum_{\bf p} 
 e^{i{\bf p}({\bf x}^{(j)}-{\bf x}^{(j')})} ~ 
e^{i\theta_a({\bf p},x_4^{(j)},\mu_j)} ~ 
e^{i\theta_{\bar a}(-{\bf p},x_4^{(j')},\mu_{j'})} ~\times~~  \nonumber\\
 \\
 \tr\{I^a W_{\vec\ell}(j~\sss_j| j'~\sss_{j'}) I^{\bar a} 
W_{\vec\ell}(j'~\sss_{j'}| j~\sss_j) \} ~ 
D^a_{|\mu_j||\mu_{j'}|}\left({\bf p};x_4^{(j)}-\delta_{\mu_j+4} ; x_4^{(j')}-\delta_{\mu_{j'}+4} \right)\,.
\nonumber
\eeqa

We introduce the propagator completely in $x$-space,
\be\label{eq:gluon_x}
\Delta^a_{\mu\nu}({\bf x};x_4,y_4)\equiv
\frac{1}{L^3} \sum_{\bf p}e^{i{\bf p}{\bf x}} ~ 
e^{i\theta_a({\bf p},x_4,\mu)} ~
e^{i\theta_{\bar a}(-{\bf p},y_4,\nu)} ~
D^a_{|\mu||\nu|}\left({\bf p}; x_4-\delta_{\mu+4} ; y_4-\delta_{\nu+4} \right)\,,
\ee
which allows us to write
\beqa
\< \tr\{ W_{\vec\ell}^{({\rm 2a})} \} \>_0 &=& \sum_{j=1}^\ell \sum_{j'=j+1}^\ell \sum_{a=1}^8
\tr\{I^a W_{\vec\ell}(j~ \sss_j| j'~\sss_{j'}) I^{\bar a} 
W_{\vec\ell}(j'~\sss_{j'}| j~ \sss_j) \} \nn[-0.5ex]
& &\la{eq:2a}\\[-0.5ex]
& &\times~~\Delta^a_{\mu_j\mu_{j'}}({\bf x}^{(j)} - {\bf x}^{(j')} ;x_4^{(j)},x_4^{(j')}).
\nn
\< \tr\{ W_{\vec\ell}^{({\rm 2b})} \} \>_0 &=& \frac{1}{2} \sum_{a=1}^8
\tr\{ I^a I^{\bar a } W_{\vec\ell}[V] \} \sum_{j=1}^\ell
\Delta^a_{\mu_j\mu_j}\left({\bf 0} ;x_4^{(j)},x_4^{(j)}\right). \la{eq:2b}
\eeqa

\subsection{Improvement}

In order to be able to reach the continuum limit with a rate proportional
to $(1/L)^2$ our observable needs to be improved. Since there are no operators
of dimension 6 with the same symmetries of ${\cal O}_{\rm spin}$, non-vanishing at one-loop
order, and with no valence quarks, the improvement amounts to compute the additional
contributions stemming from the volume and boundary counter-terms in the action.
The volume term is proportional to $c_{\rm sw}$, whose tree-level expression, $c_{\rm sw}^{(0)}=1$, enters 
our observable at one-loop order. It is taken into account directly in the quark propagator.
The only boundary term needed is proportional to the one-loop expression of $\ct$ \cite{alpha:SU3,pert:1loop}.
The corresponding counterterm can be expressed as 
\be
\< \tr\{W_{\vec\ell}^{(1)}\}  S^{(1)}_{\rm tot,b} \>_0
\ee
with $S^{(1)}_{\rm tot,b}$ given in eq.~(5.130) of (K). The explicit expression reads
\be\label{eq:impr_term}
\< \tr\{W_{\vec\ell}^{(1)}\}  S^{(1)}_{\rm tot,b} \>_0=\frac{2}{\sqrt{3}}
c_{\rm t}^{(1)}[\sin(2\gamma)+\sin(\gamma)]\tr\{I^8W_{\vec\ell}[V]\}\sum_{k=1}^3M_{\vec\ell,k}\,,
\ee
with
\bea
& &\hspace{-1.5cm}M_{\vec{\ell},k}=\sum_{j=1}^{\ell}{\rm sign}(\mu_j)\biggr(\delta_{|\mu_j|-4}+(1-\delta_{|\mu_j|-4})
{\rm e}^{-i\phi_8(x_4^{(j)})/2}\biggr)\label{eq:Mlk}\\
             & &\times~~
\biggr(D^8_{k|\mu_j|}(\mathbf{0},1,x_4^{(j)}-\delta_{\mu_j+4})-D^8_{k|\mu_j|}
(\mathbf{0},T-1,x_4^{(j)}-\delta_{\mu_j+4}) \biggr)\nonumber\,,
\eea
and $\gamma=\pi/3LT$ once the ``point A'' has been chosen. The contribution (\ref{eq:impr_term})
vanishes for the Polyakov loop ${\cal P}_3(x)|_{x_4=T/2}^{\phantom{\dagger}}$ without operator insertion.

\subsection{Summary\\}
The expectation value of the Wilson loop at one-loop order is given by
\bea
& &\hspace{-1.5cm}\< \tr\{ W_{\vec\ell} \} \> = W_{\vec\ell}[V]+ g_0^2\left( 
\< \tr\{ W_{\vec\ell}^{({\rm 2a})} \} \>_0 + 
\< \tr\{ W_{\vec\ell}^{({\rm 2b})} \} \>_0\right.\\
& &\hspace{3cm}\left. -~ \< \tr\{W_{\vec\ell}^{(1)}\}  S^{(1)}\>_0 - 
\< \tr\{W_{\vec\ell}^{(1)}\}  S^{(1)}_{\rm tot,b} \>_0 \right)\,,\nonumber
\eea
where the one-loop terms are given by eqs.~(\ref{eq:2a},\ref{eq:2b},\ref{eq:11}) and 
the improvement term by \eq{eq:impr_term}.

\subsection{Implementation in {MATLAB}}

For our perturbative computations we decided to use MATLAB
in order to combine comfortable programming, robustness of the libraries 
and acceptable speed for the involved observables and lattices. 

In presence of a non-vanishing background field, a simple analytical expression 
for the ghost, gluon and quark propagators is not available. They are computed
by exploiting the recursive techniques presented in \cite{pert:2loop_SU2,pert:2loop_fin}.
The gluon propagator is the most time consuming computation. Its Fourier transformed 
expression (\ref{eq:gluon_x}) is calculated by summing only over a reduced set 
of momenta, which saves a factor of 6 (asymptotically on large lattices) in 
computing time.

The tadpole loops are observable independent, and they are computed and stored.
We use the formulae of \cite{Weisz:int_csw}, where the symmetries of vertices and propagators
are fully exploited. Then the contributions (\ref{eq:11}) are computable with an effort
negligible in comparison to the loops.

The improvement counterterm involves only the zero momentum gluon propagator and the
trace of the product of diagonal matrices;
it is computationally cheap in comparison to the rest. 

In order to give an idea of the computational cost,
for $L=48$ the computation of all diagrams
and improvement counter-terms for the Polyakov loop with insertion
of the clover leaf operator has been carried out in 2 weeks on a PC,
equipped with a single processor Intel Pentium 4 with 2.6 GHz.
The scaling can be approximated with a polynomial in $L$,
and is asymptotically dominated by the highest power, i.e.~$L^5$.

\begin{figure}[!p]
\begin{center}
\begin{tabular}{cc}
\includegraphics[height=5.8cm,width=6.8cm]{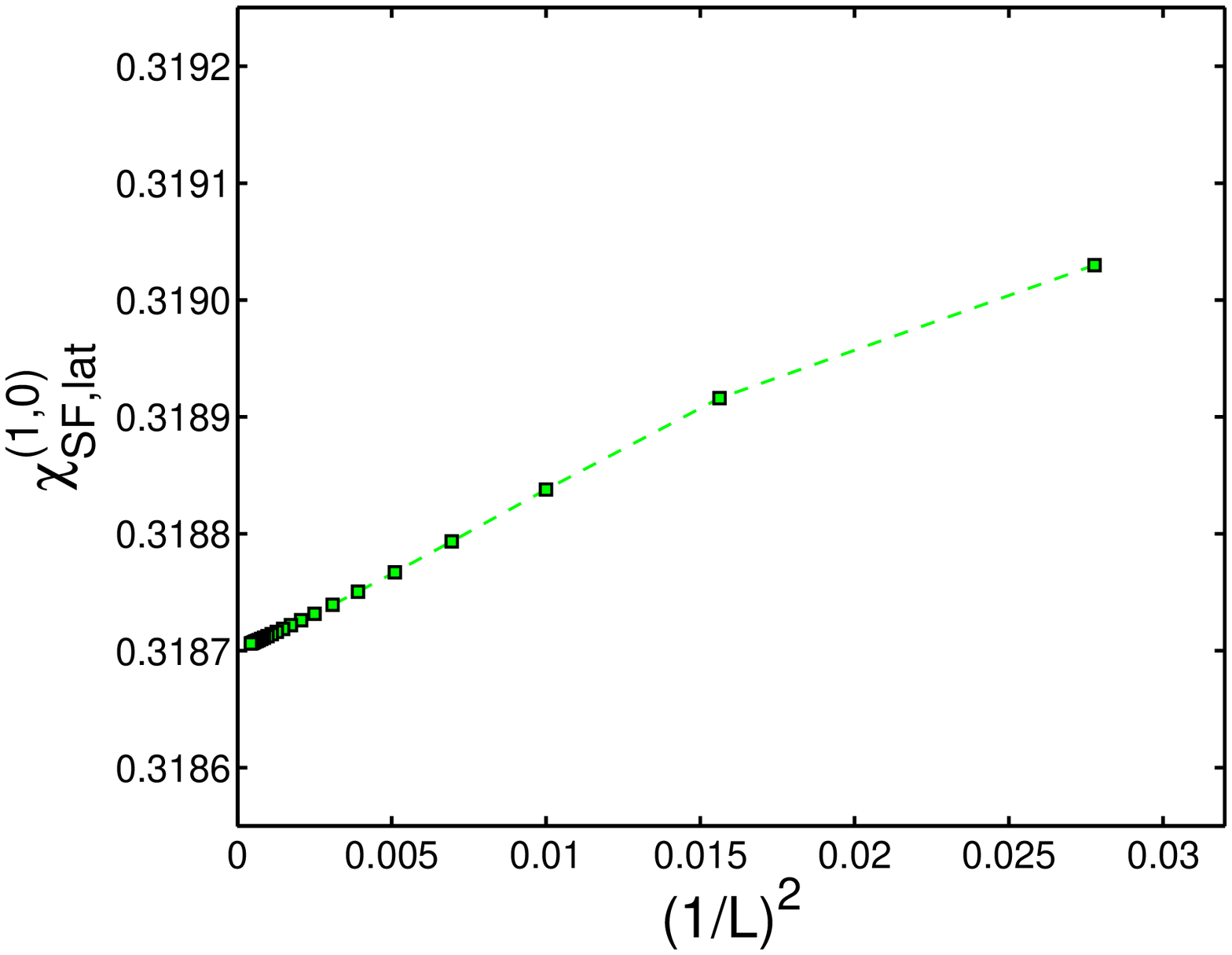} &
\includegraphics[height=5.8cm,width=6.8cm]{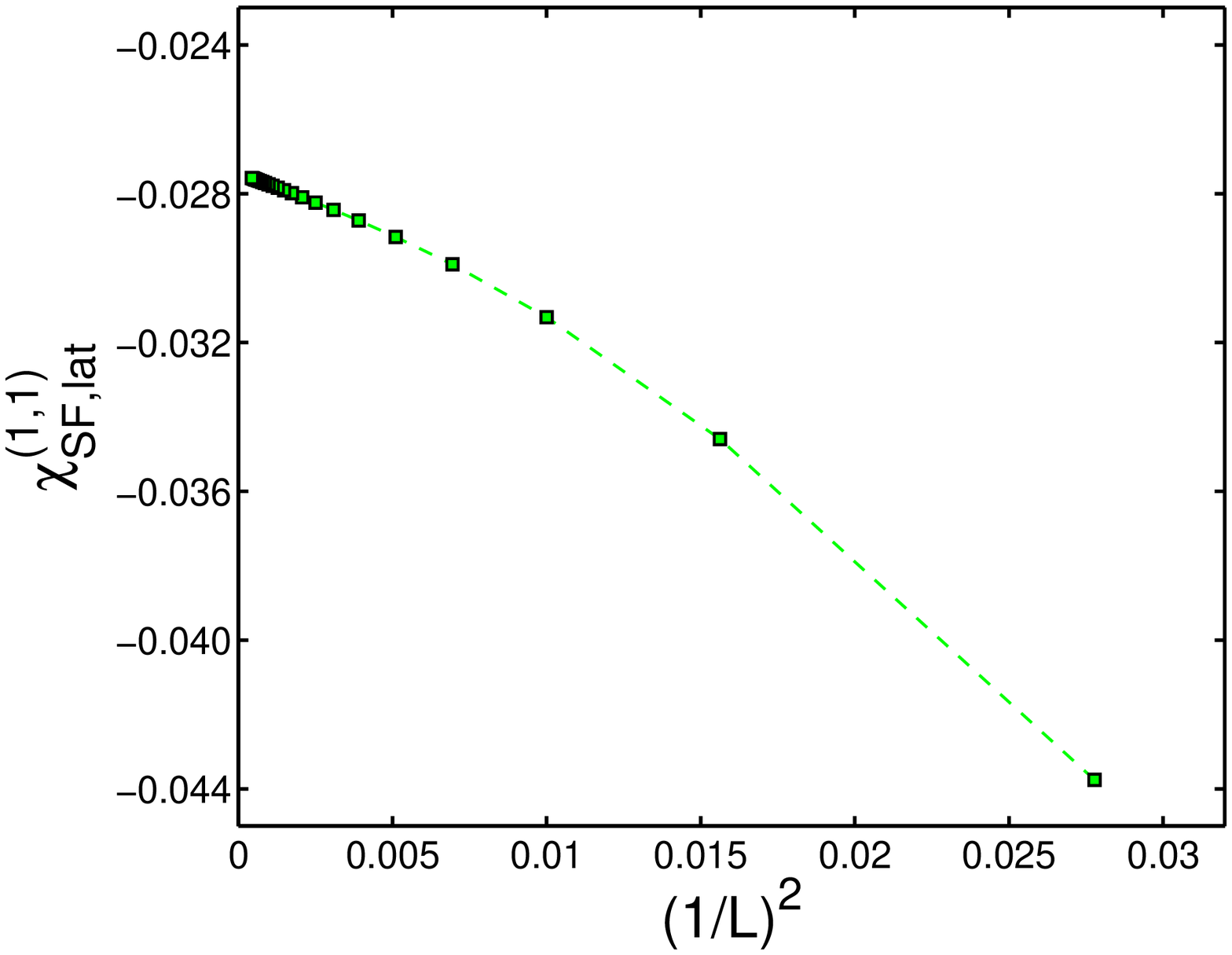} \qquad \\
\end{tabular}
\end{center}
\caption[]{\footnotesize
{
One-loop connection between the SF scheme and the lat scheme.
}}\label{fig:Chi_spin}
\vspace{0.7cm}
\end{figure}

\TABLEP{
\begin{tabular}{ccc}
\hline \\[-1.75ex]
$L$   & $\chi_{\rm SF,lat}^{(1,0)}$ & 
$\chi_{\rm SF,lat}^{(1,1)}$   \\[1ex]
\hline \\[-1.75ex]
4  &   0.319032402694607  & -0.071383603501862 \\
6  &   0.319029788544510  & -0.043758342410953 \\
8  &   0.318915972669536  & -0.034594298272846 \\
10 &   0.318837839980287  & -0.031316158100235 \\
12 &   0.318793467352611  & -0.029898520735474 \\
14 &   0.318767160420896  & -0.029160440246041 \\
16 &   0.318750540373758  & -0.028720855529176 \\
18 &   0.318739443194014  & -0.028434643639671 \\
20 &   0.318731691659922  & -0.028236618214183 \\
22 &   0.318726074540605  & -0.028093427502346 \\
24 &   0.318721879237185  & -0.027986322197915 \\
26 &   0.318718665931215  & -0.027904013713480 \\
28 &   0.318716151870336  & -0.027839341836381 \\
30 &   0.318714148841999  & -0.027787573356721 \\
32 &   0.318712527751079  & -0.027745471721788 \\
34 &   0.318711197727617  & -0.027710760078004 \\
36 &   0.318710093332871  & -0.027681797428228 \\
38 &   0.318709166494586  & -0.027657376001968 \\
40 &   0.318708381269461  & -0.027636590425178 \\
42 &   0.318707710343347  & -0.027618750920289 \\
44 &   0.318707132670872  & -0.027603324328397 \\
46 &   0.318706631831595  & -0.027589893181705 \\
48 &   0.318706194851218  & -0.027578126762064 \\
 & & \\[-1.75ex]
\hline
\end{tabular}
\caption{\footnotesize 
Results for the connection between the SF and the lattice MS schemes.}\label{t:chi_spin}
}

\subsection{Checks}\label{sect:av_plaqi}
Of all checks we did to confirm the correctness of our code, we briefly
report about two of them, which may be of interest in other
applications.

As observed in \cite{Creutz:1984mg}, 
the expectation value of the gauge action can be evaluated 
economically by taking the logarithmic derivative
of the partition function with respect to $\beta=6/g_0^2$:
\beqa
\frac{1}{3}\< {\rm Re~\tr}\{1-P_{\mu\nu}(x)\}\> &=& 
\frac{1}{2\beta}\cdot 
\frac{\textrm{ Nb. of propagating gluons}}{\textrm{ Nb. of un-oriented plaquettes}}
+ {\rm O}(1/\beta^2)
\\ &=& \frac{1}{2\beta}\cdot \frac{8[L^3(4T-3) - L^3(T-1)]-\nu}{3L^3(2T-1)} 
    + {\rm O}(1/\beta^2).
\eeqa
The  term $\nu$ arises from the gauge degrees of freedom that are
constant in space and live on the lower temporal boundary
(there is no extra gauge degree of freedom associated with the 
boundary $x_4=T$ because global symmetries are not to be gauge-fixed): 
\be
\nu =  \left\{ \begin{array}{l@{\qquad}l}
{\rm dim}({su}(3))=8 & \textrm {with boundary links set to unity} \\
{\rm rank}({su}(3))= 2 & \textrm {with non-trivial Abelian boundary field}
\end{array}
\right.
\ee 
In the first case, all 8 zero momentum gluons at the lower temporal boundary
obey Dirichlet boundary conditions. They are associated with spatially constant
modes and are therefore not propagating  degrees of freedom.
With a non-trivial Abelian background field, 
only two of the gluon fields 
obey the Dirichlet boundary conditions at the lower boundary, 
and are associated with spatially constant diagonal modes;
the others are propagating modes. 
We checked that our program reproduces this result.

A further successful check, which we do not report in detail, 
consists in comparing the perturbative results
for the plaquette and the Polyakov loops, with and 
without insertion of the clover leaf operator,
to the corresponding non-perturbative (quenched) computations. 
The latter are performed 
at small bare couplings, $0.015\leq g_0^2 \leq 0.06$, 
setting all needed improvement coefficients
to their tree-level values. In all cases $L=T=4$.

\subsection{The Polyakov loop and chromo-magnetic operator}

\subsubsection{Tree-level computation}
As far as the gauge boundary values and the induced background field are concerned,
we follow \cite{alpha:SU3}, and work
with the boundary fields defined by eqs. (\ref{e:Ubound},\ref{e:CC}).
The numerator and denominator on the r.h.s.~of the renormalization condition (\ref{e:zspindef3})
assume the compact form
\beqa
L^2  \< \tr (\pol_3(x)E_{1}(x)) \>_{\gounder=\,0} 
&=&L^2\sum_{m=1}^{3}\exp\{\frac{i}{L}[x_4 \angleprime_m
+(L-x_4)\angle_m]\}\nonumber\\[-1.5ex]
&&\label{e:PEtree} \\[-1.5ex]
&& ~~~~~~~~~~~{\scriptsize \times}\sin\left[\frac{1}{L^2}\left(\angleprime_m-\angle_m\right)\right]. \nonumber \\
\< \tr \pol_3(x)\>_{\gounder=\,0} 
&=&  \sum_{m=1}^{3}{\rm exp}\left\{\frac{i}{L}[x_4 \angleprime_m + (L-x_4)\angle_m]\right\}
\label{e:P3tree} 
\eeqa

\subsubsection{One-loop order}

\begin{figure}[t]
\begin{center}
\begin{tabular}{cc}
\includegraphics[scale=0.4]{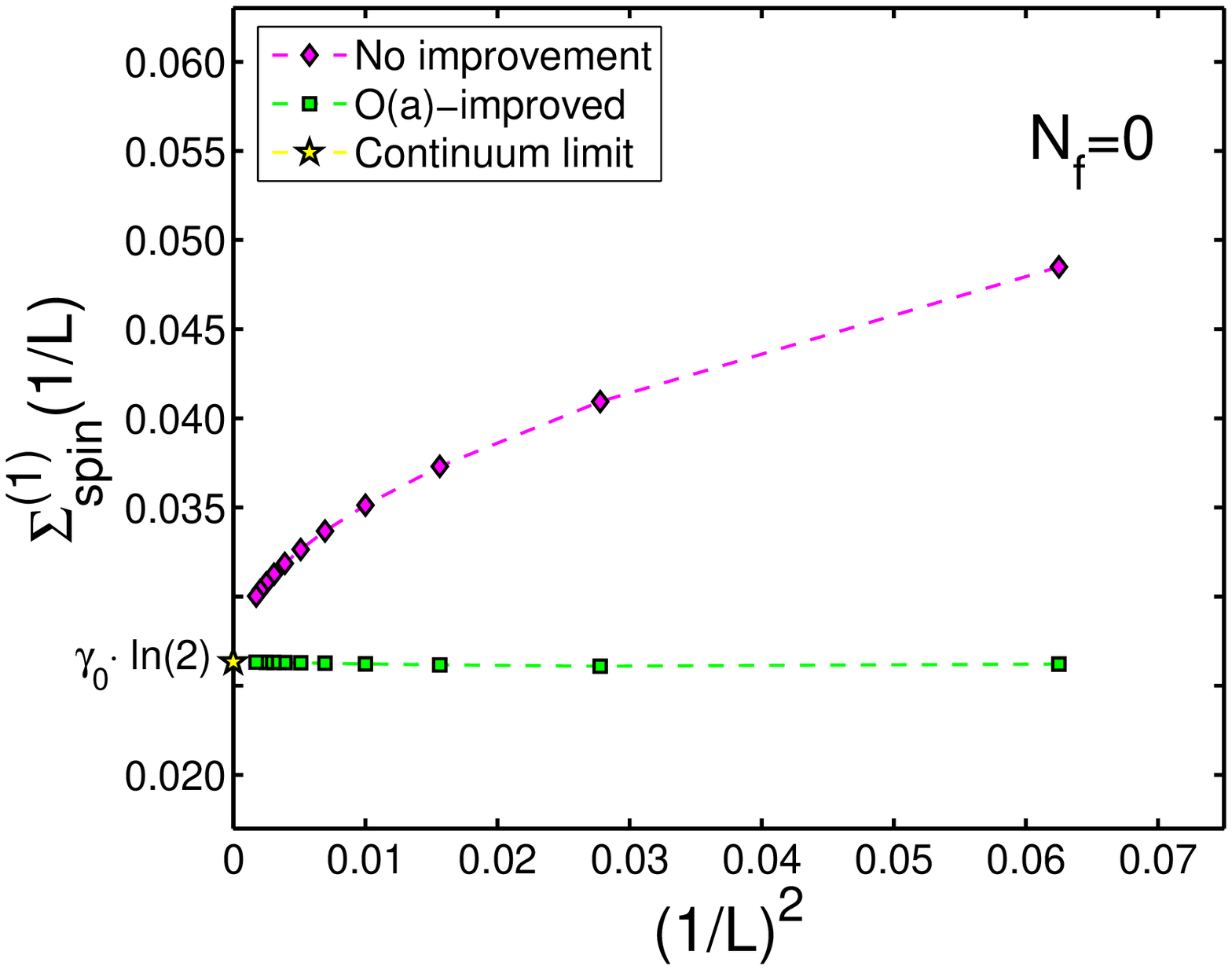} &
\includegraphics[scale=0.4]{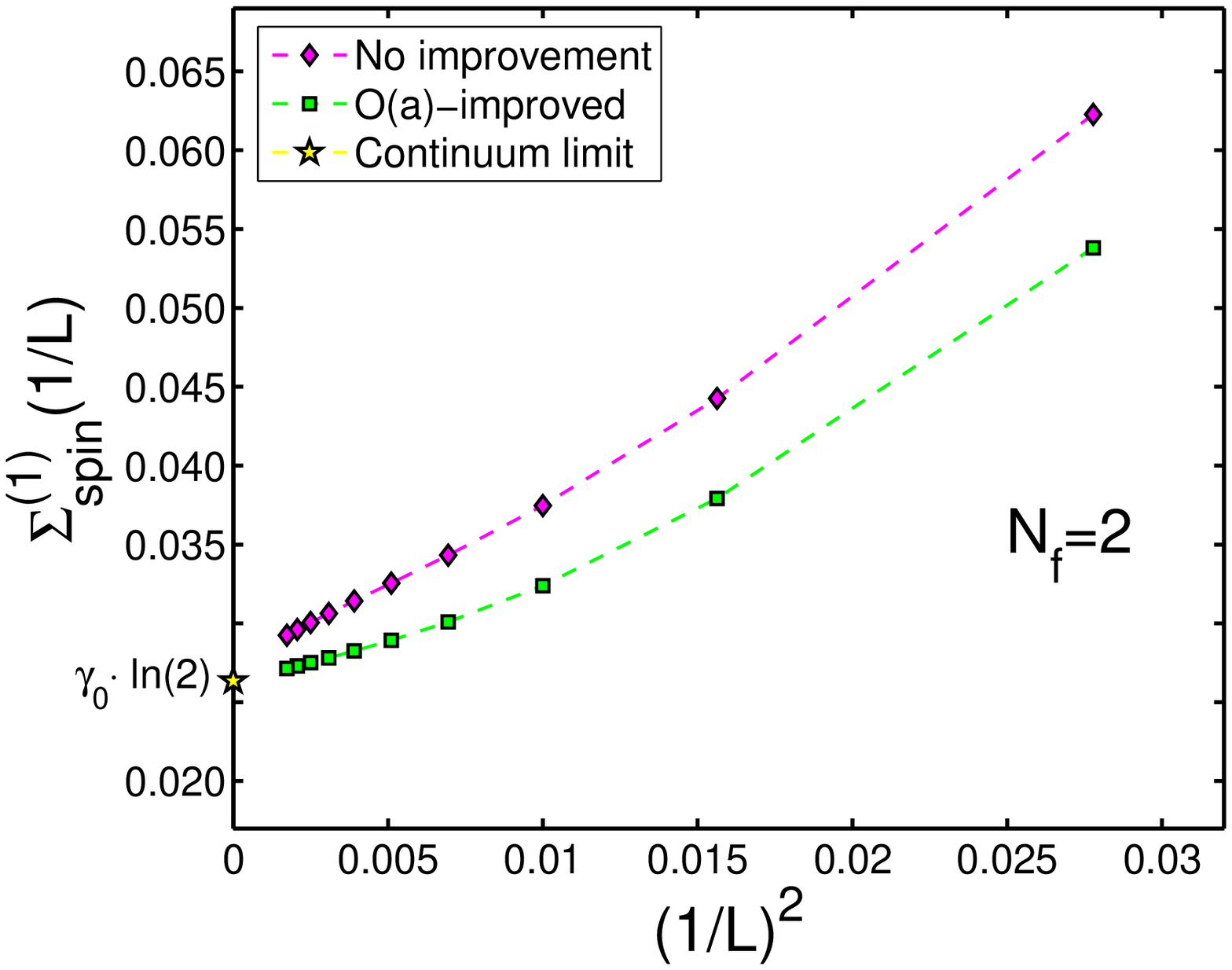} \qquad \\
\end{tabular}
\end{center}
\caption[]{\footnotesize
{One-loop contribution to the step scaling function of the chromo-magnetic operator
in the $N_{\rm f}=0$ (left) and $N_{\rm f}=2$ (right) cases.}}\label{fig:Sigma_spin}
\end{figure}

The one-loop contribution to the lattice 
step scaling function $\Sigma_{\rm spin}(u,1/L)=1+\Sigma_{\rm spin}^{(1)}(1/L)u+{\rm O}(u^2)$, 
defined in \eq{eq:Sigma_spin}, is computed for $N_{\rm f}=0$ and 
$N_{\rm f}=2$ giving the results represented in \Fig{fig:Sigma_spin}. There the effect 
of the $\Oa$-improvement is evident, especially in the $N_{\rm f}=0$ case. In both the unimproved
and improved cases the continuum limit is consistent with the prediction $\gamma_0 \ln(2)$.

These results enter the computation of the connection 
between the SF scheme and the lat scheme, as shown in \Fig{fig:Chi_spin}.
The one-loop connection factor $\chi_{\rm SF,lat}^{(1)}$ is obtained from the one-loop
contribution $Z_{\rm spin}^{\rm SF,(1)}$ to the renormalization factor (\ref{e:zspindef3})
by subtracting the logarithmic divergent part
\be\label{eq:def_chi_1}
\chi_{\rm SF,lat}^{(1)}(L)=Z_{\rm spin}^{\rm SF,(1)}(L)-\gamma_0\ln(L)\,.
\ee
We decompose $\chi_{\rm SF,lat}^{(1)}$ according to its $N_{\rm f}$-dependence
\be\label{eq:Chi_SF_lat_1}
\chi_{\rm SF,lat}^{(1)}=\chi_{\rm SF,lat}^{(1,0)}+N_{\rm f}\chi_{\rm SF,lat}^{(1,1)}\,.
\ee
The quarks are massless, implemented at this order in perturbation
theory by $m_0=0$ and the angle $\theta=-\pi/3$.
The full listing of results is shown in \tab{t:chi_spin}, where all numbers are
given with 15 digits for readability, although the last two or three may be insignificant.

The continuum limits, including the estimate of the associated uncertainties, are 
performed according to the method described in \cite{pert:2loop_fin},
with MATLAB routines provided by Ulli Wolff.
We have verified that the roundoff errors 
as well as the errors in $\ct$ quoted in \cite{alpha:SU3,pert:1loop} 
are negligible
compared to the systematic uncertainties of the extrapolation. 
The final result of the continuum limit extrapolations is expressed in \eq{eq:Chi_SF_lat_res}. 

\section{Monte Carlo simulations \label{s:MC}}

In our measurements of observables we fully exploit
translational and axis exchange invariance. 
The ensemble of gauge configurations
is generated by means of the ``hybrid over-relaxation'' algorithm 
with lexicographically ordered sweeps
(see e.g.~\cite{deDivitiis:1995yp} for the exact implementation). 
The basic update consists
of 1 heat-bath update sweep \cite{Cabibbo:1982zn,Fabricius:1984wp,Kennedy:1985nu}, 
followed by $N_{\rm OR}$ over-relaxation
sweeps \cite{HOR1}.  The update is iterated $N_{\rm UP}=2$ times between measurements
and the parameter $N_{\rm OR}$ 
varies from a minimum of $3$, for $L/a=6$, to a maximum of $10$, for $L/a=24$. This 
guarantees to have short integrated autocorrelation times for our observables, 
while the computing time
spent for the update does not exceed the one required for the measurements. 
Still there can be very slow modes in the system as will be discussed  
at the end of the appendix.

In the case of the Eichten-Hill action, the gauge links 
building up the Polyakov loop, but not the inserted clover leaf operator,
are evaluated by a 10-hit multi-hit procedure \cite{PPR}, 
where each hit consists of a heat-bath 
update of the above type. With this variance reduction, the statistical
precision is similar to the one of the HYP2 action. On the largest lattice
($L/a=24$) 
we could obtain an around 1\% precision in $\zmagsf(L)$ with 
30k measurements at $\beta\approx11$ and with 100k measurements
at $\beta\approx 7.2$. 
However, at $\beta\approx 6.8$, where the length of
of the Polyakov loop amounts to $0.7\,\fm$, it became 
very costly to reach even a 2\% precision. We tried various 
ways to reduce the variance, in particular different versions
of multilevel algorithms inspired by \cite{algo:LW1}, but did
not succeed in finding significant gains. We then changed the discretization 
of $\Ospin$, replacing the links in $\hat{F}_{\mu\nu}$ by HYP2 links.
(This is indicated throughout the paper as the discretization with HYP2 action 
and HYP2 operator $\hat{F}_{\mu\nu}$.)
The resulting increase in precision allowed to obtain
the last entry in \tab{t:Sigma} with 250k measurements.

\begin{figure}[pt]
\begin{center}
\begin{tabular}{cc}
\includegraphics[scale=0.35]{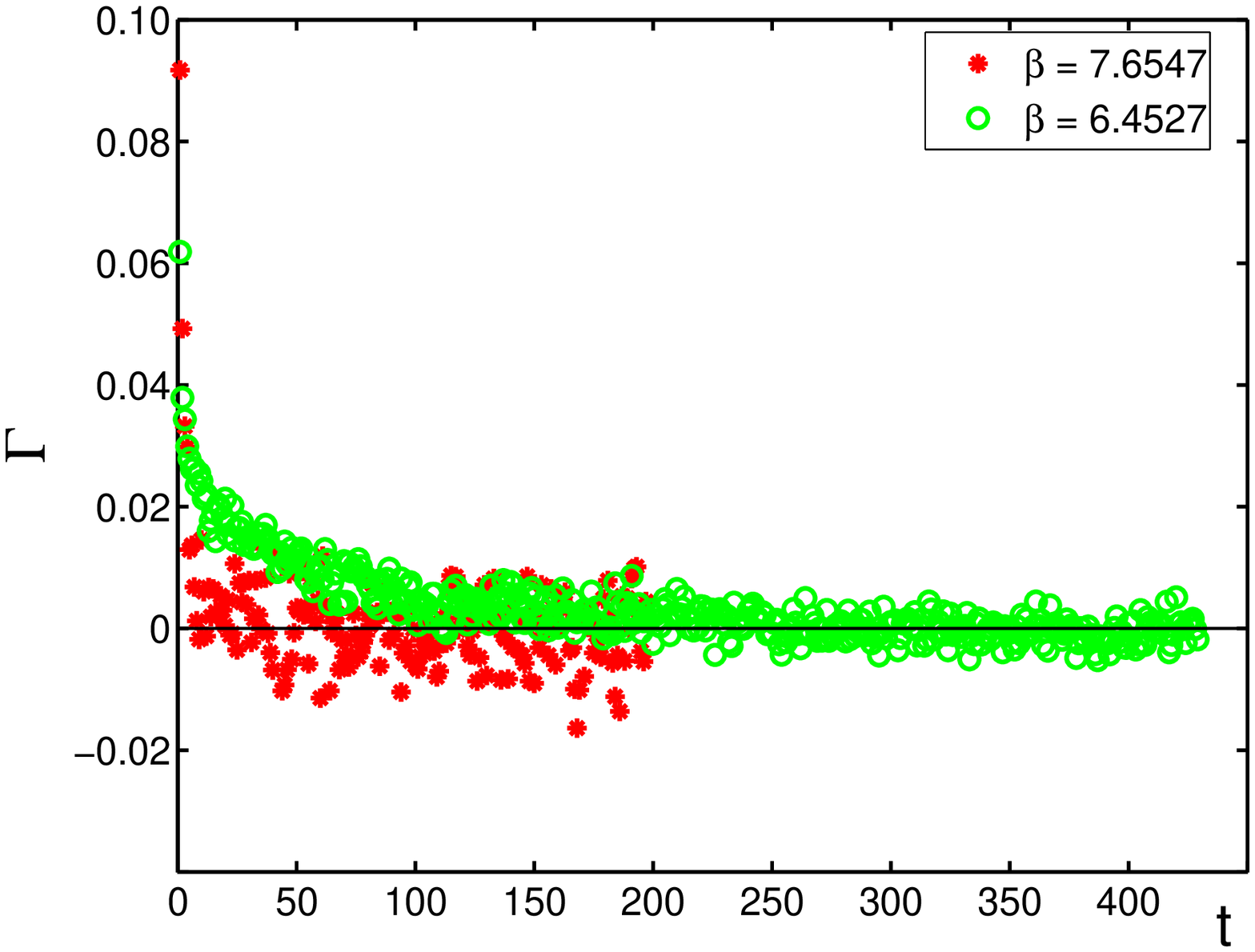} &
\vspace{-0.1cm}\includegraphics[scale=0.35]{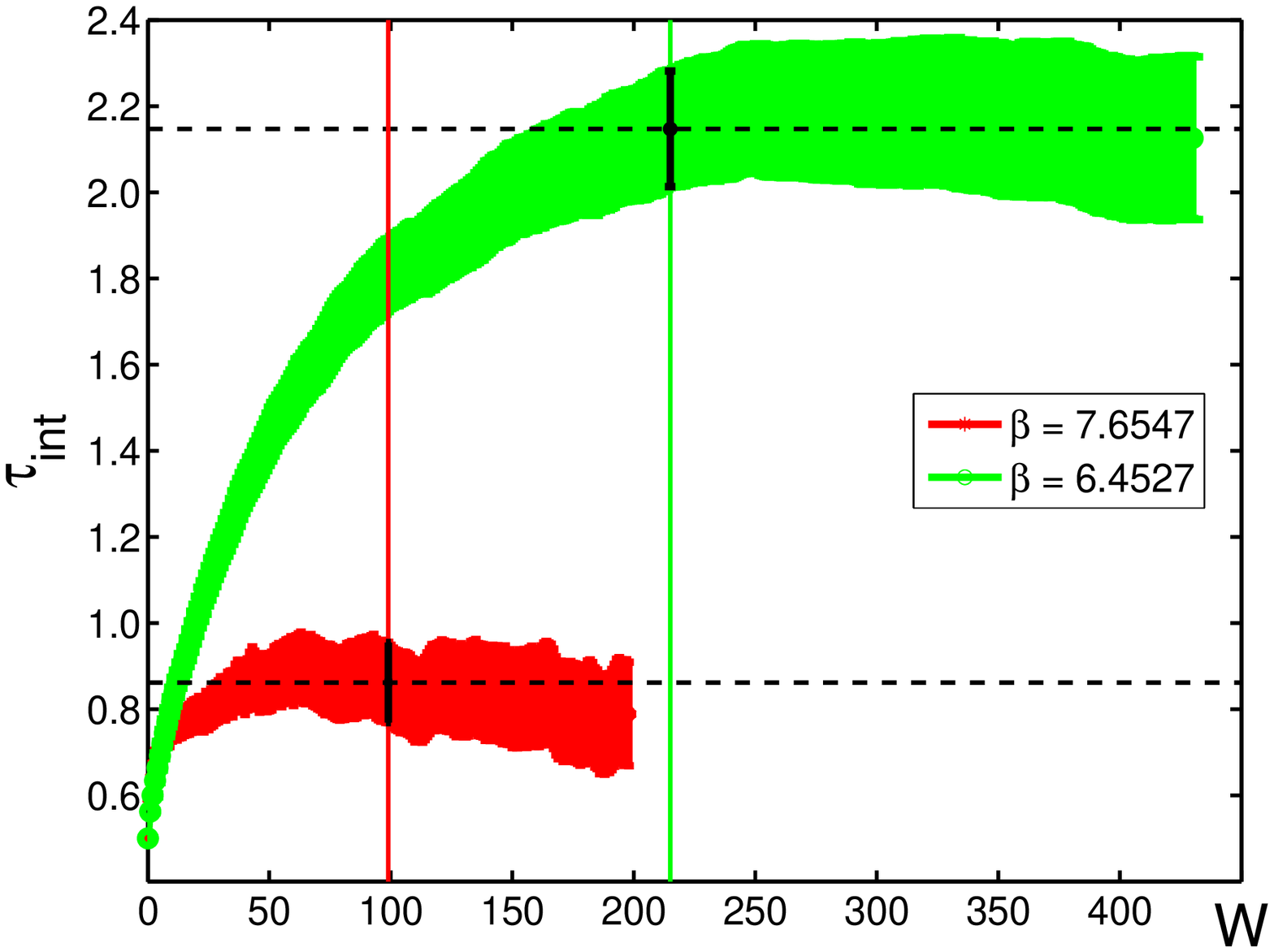} \qquad \\
\end{tabular}
\end{center}
\caption[]{\footnotesize Autocorrelation pattern of $Z_{\rm spin}(2L)$ 
for two quenched simulations with $L/a=16$.
{Left: Normalized autocorrelation function plotted vs. the separation of
measurements $t$.
Right: Integrated autocorrelation time $\tau_{\rm int}$ vs. the summation
window~${\rm W}$.
With $\beta=7.6547$ we have $\gbsq(L)=1.8811$ (red asterisks), while for
$\beta=6.4527$ we have $\gbsq(L)=3.480$ (green circles). The vertical lines correspond
to the optimal values of ${\rm W}$ computed according to \cite{Wolff:2003sm}.}}\label{f:Gamma_plot}
\end{figure}


\TABLEP{
\begin{tabular}{ccccclll}
\hline \\[-1.75ex]
$\gbar^2(L)$ & $\beta$ & $L / a$   & action & $\hat{F}_{\mu\nu}$ & $\zmagsf(L)$  
             & $\zmagsf(2L)$ & $\Sigmamag$ \\[1ex]
\hline \\[-1.75ex]
0.8873(5) & 10.7503& 6  & & &1.3188(46) &1.3504(50) &1.0239(52) \\
0.8873(10)& 11.0000& 8  & & &1.3218(22) &1.3532(51) &1.0238(42) \\
0.8873(30)& 11.3384& 12 & & &1.3268(30) &1.3580(69) &1.0235(57) \\
\hline
0.9944(7) & 10.0500& 6  &EH & &1.3651(44) &1.3905(76) &1.0186(64)  \\
0.9944(13)& 10.3000& 8  &EH & &1.3514(52) &1.3924(88) &1.0303(76)  \\
0.9944(30)& 10.6086& 12 &EH & &1.3608(53) &1.384(12)  &1.0171(96)  \\
\hline
1.2430(13)& 8.8997 & 6  & & &1.4336(44) &1.4839(65) &1.0351(55) \\
1.2430(14)& 9.1544 & 8  & & &1.4278(30) &1.4803(56) &1.0367(45) \\
1.2430(35)& 9.5202 & 12 & & &1.4349(27) &1.474(10)  &1.0275(74) \\
\hline
1.3293(12)& 8.6129 & 6  &EH & &1.4727(57) &1.5116(77)&1.0264(66) \\
1.3293(21)& 8.8500 & 8  &EH & &1.4664(71) &1.503(12) &1.0248(95) \\
1.3293(60)& 9.1859 & 12 &EH & &1.4528(65) &1.517(13) &1.0438(99) \\
\hline
1.5553(15)& 7.9993 & 6  & & &1.5302(61) &1.6036(48) &1.0480(53) \\
1.5553(24)& 8.2500 & 8  & & &1.5229(46) &1.5999(78) &1.0506(60) \\
1.5553(70)& 8.5985 & 12 & & &1.5191(27) &1.579(11)  &1.0394(77) \\
\hline
1.8811(22)& 7.4082 & 6  & & &1.6193(74) &1.7208(65) &1.0627(63) \\
1.8811(28)& 7.6547 & 8  & & &1.6100(55) &1.717(13)  &1.0664(91) \\
1.8811(38)& 7.9993 & 12 & & &1.6015(42) &1.718(14)  &1.0726(94) \\
\hline
2.1000(39)& 7.1214 & 6  & & &1.6563(87) &1.792(12) &1.0819(92) \\
2.1000(45)& 7.3632 & 8  & & &1.652(11)  &1.773(11) &1.0732(96) \\
2.1000(80)& 7.6985 & 12 & & &1.6577(91) &1.751(17) &1.056(12) \\
\hline
2.4484(37)& 6.7807 & 6  & & &1.7618(88) &1.900(13) &1.0786(90) \\
2.4484(45)& 7.0197 & 8  & & &1.7371(86) &1.898(17) &1.092(11) \\
2.4484(80)& 7.3551 & 12 & & &1.7141(89) &1.855(17) &1.082(12) \\
\hline
2.770(7)  & 6.5512 & 6  & & &1.8317(75) &2.085(19) &1.138(11) \\
2.770(7)  & 6.7860 & 8  & & &1.8067(94) &2.044(17) &1.131(11) \\
2.770(11) & 7.1190 & 12 & & &1.7975(90) &2.000(23) &1.113(14) \\
\hline
2.770(7)  & 6.5512 & 6  & &HYP2 &1.3659(36) &1.501(12) &1.0986(91) \\
2.770(7)  & 6.7860 & 8  & &HYP2 &1.3643(24) &1.505(15) &1.103(11) \\
2.770(11) & 7.1190 & 12 & &HYP2 &1.3668(63) &1.490(14) &1.090(11) \\
\hline
3.480(8)  & 6.2204 & 6  & &HYP2 &1.4329(53) &2.070(29) &1.444(21) \\
3.480(14) & 6.4527 & 8  & &HYP2 &1.4350(89) &1.975(31) &1.376(23)  \\
3.480(39) & 6.7750 & 12 & &HYP2 &1.4465(61) &1.937(41) &1.339(29) \\
\hline
3.480(8)  & 6.257  & 6  & & &1.9864(75) &2.714(32) &1.366(17) \\
3.480(8)  & 6.476  & 8  & & &1.9492(75) &2.608(33) &1.338(18) \\
\hline
3.480(8)  & 6.257  & 6  & &HYP2 &1.4297(47) &1.979(26) &1.384(19) \\
3.480(8)  & 6.476  & 8  & &HYP2 &1.4262(29) &1.961(25) &1.375(17) \\
3.480(9)  & 6.799  & 12 & &HYP2 &1.4280(33) &1.864(37) &1.305(26) \\
          &        &    & &      &           &          &          \\[-1.75ex]
\hline
\end{tabular}
\caption{Raw simulation results. The standard discretization is the 
HYP2 action of \cite{stat:actpaper} and the clover leaf 
operator $\hat{F}_{\mu\nu}$. Deviations
from this rule are indicated. The improvement coefficient $\ct$ is set
to its 1-loop value, except for the last five lines, where 2-loop
precision is used. The renormalized coupling is reproduced from
\cite{mbar:pap1}.
}\label{t:Sigma}
}

As a check that the
change of discretization does not introduce unwantedly large $a^2$ effects,
we also repeated the computation of $\sigma(2.77)$ this way. \Fig{f:Sigma}
nicely confirms the expected universality and $a^2$ effects actually 
turn out to be reduced! 
We point out that with the Eichten-Hill action
the cutoff effects can be directly compared 
to the expectations of perturbation theory. 
For the investigated couplings, $\gbsq=0.9944,1.3293$, the agreement is very good.

The large amount of statistically independent measurements needed 
imposes strong limitations to the application of this method to the 
theory with 
dynamical fermions. The raw simulation results are reported in \Tab{t:Sigma}. 

We finally add a remark on error estimates and autocorrelations in 
our simulations.
The autocorrelation function $\Gamma$ of $Z_{\rm spin}$, defined as in
\cite{Wolff:2003sm}, falls very quickly for all our mesurements. 
At all but the largest coupling the integrated autocorrelation
time is then easily estimated. However for $L\approx0.7\,\fm$, 
we observe that $\Gamma$ shows a long tail before approaching
zero. This pattern is absent in the smaller volumes, and shows little sensitivity
to changes of $N_{\rm OR}$. \Fig{f:Gamma_plot} (left) plots
the $\Gamma$-function obtained from runs 
with $(L/a,N_{\rm UP},N_{\rm OR})=(16,2,4)$, 
for two different physical volumes. 
\Fig{f:Gamma_plot} (right) shows that for $\beta=7.6547$ the integrated autocorrelation 
time is quite short, i.e.~$\tau_{\rm int}=0.86(9)$, whereas for $\beta=6.4527$ the 
tail mentioned above
leads to $\tau_{\rm int}=2.15(13)$. The latter translates into an increase of the error 
by a factor of two, in comparison to the case where no 
correlation is present (i.e. $\tau_{\rm int}=0.5$).

\end{appendix}


\begin{thebibliography}{10}

\bibitem{stat:eichhill1}
E.~Eichten and B.~Hill, {\it An effective field theory for the calculation of
  matrix elements involving heavy quarks},  {\em Phys. Lett.} {\bf B234} (1990)
  511.

\bibitem{hqet:cont3}
B.~Grinstein, {\it The static quark effective theory},  {\em Nucl. Phys.} {\bf
  B339} (1990) 253--268.

\bibitem{hqet:cont4}
H.~Georgi, {\it An effective field theory for heavy quarks at low energies},
  {\em Phys. Lett.} {\bf B240} (1990) 447--450.

\bibitem{stat:bbstar}
M.~Bochicchio, G.~Martinelli, C.~R. Allton, C.~T. Sachrajda, and D.~B.
  Carpenter, {\it Heavy quark spectroscopy on the lattice},  {\em Nucl. Phys.}
  {\bf B372} (1992) 403--420.

\bibitem{hqet:spinsplitt_gms}
V.~Gimenez, G.~Martinelli, and C.~T. Sachrajda, {\it A high-statistics lattice
  calculation of $\lambda_1$ and $\lambda_2$ in the {B}-meson},  {\em Nucl.
  Phys.} {\bf B486} (1997) 227--244,
  [\href{http://xxx.lanl.gov/abs/hep-lat/9607055}{{\tt hep-lat/9607055}}].

\bibitem{hqet:spinsplitt_jlqcd}
{\bf JLQCD} Collaboration, S.~Aoki {\em et.~al.}, {\it Heavy quark expansion
  parameters from lattice nrqcd},  {\em Phys. Rev.} {\bf D69} (2004) 094512,
  [\href{http://xxx.lanl.gov/abs/hep-lat/0305024}{{\tt hep-lat/0305024}}].

\bibitem{stat:eichhill_1m}
E.~Eichten and B.~Hill, {\it Static effective field theory: $1/m$ corrections},
   {\em Phys. Lett.} {\bf B243} (1990) 427--431.

\bibitem{stat:ospin_flynn}
J.~M. Flynn and B.~R. Hill, {\it {B} - {B}$^*$ splitting: A test of heavy quark
  methods},  {\em Phys. Lett.} {\bf B264} (1991) 173--177.

\bibitem{alpha:sigma}
M.~{L\"uscher}, P.~Weisz, and U.~Wolff, {\it A numerical method to compute the
  running coupling in asymptotically free theories},  {\em Nucl. Phys.} {\bf
  B359} (1991) 221--243.

\bibitem{alpha:letter}
{\bf ALPHA} Collaboration, A.~Bode {\em et.~al.}, {\it First results on the
  running coupling in {QCD} with two massless flavors},  {\em Phys. Lett.} {\bf
  B515} (2001) 49--56, [\href{http://xxx.lanl.gov/abs/hep-lat/0105003}{{\tt
  hep-lat/0105003}}].

\bibitem{mbar:pap1}
{\bf ALPHA} Collaboration, S.~Capitani, M.~{L\"uscher}, R.~Sommer, and
  H.~Wittig, {\it Non-perturbative quark mass renormalization in quenched
  lattice {QCD}},  {\em Nucl. Phys.} {\bf B544} (1999) 669,
  [\href{http://xxx.lanl.gov/abs/hep-lat/9810063}{{\tt hep-lat/9810063}}].

\bibitem{hqet:pap1}
{\bf ALPHA} Collaboration, J.~Heitger and R.~Sommer, {\it Non-perturbative
  heavy quark effective theory},  {\em JHEP} {\bf 02} (2004) 022,
  [\href{http://xxx.lanl.gov/abs/hep-lat/0310035}{{\tt hep-lat/0310035}}].

\bibitem{hqet:pap4}
M.~Della~Morte, N.~Garron, M.~Papinutto, and R.~Sommer, {\it Heavy quark
  effective theory computation of the mass of the bottom quark},  {\em JHEP}
  {\bf 01} (2007) 007, [\href{http://xxx.lanl.gov/abs/hep-ph/0609294}{{\tt
  hep-ph/0609294}}].

\bibitem{mb:steinhkuehn2}
J.~H. Kuhn, M.~Steinhauser, and C.~Sturm, {\it Heavy quark masses from sum
  rules in four-loop approximation},
  \href{http://xxx.lanl.gov/abs/hep-ph/0702103}{{\tt hep-ph/0702103}}.

\bibitem{mbar:charm1}
{\bf ALPHA} Collaboration, J.~Rolf and S.~Sint, {\it A precise determination of
  the charm quark's mass in quenched {QCD}},  {\em JHEP} {\bf 12} (2002) 007,
  [\href{http://xxx.lanl.gov/abs/hep-ph/0209255}{{\tt hep-ph/0209255}}].

\bibitem{lat06:damiano}
D.~Guazzini, R.~Sommer, and N.~Tantalo, {\it $m_b$ and $f_{B_s}$ from a
  combination of hqet and qcd},  {\em PoS} {\bf LAT2006} (2006) 084,
  [\href{http://xxx.lanl.gov/abs/hep-lat/0609065}{{\tt hep-lat/0609065}}].

\bibitem{zastat:pap1}
{\bf ALPHA} Collaboration, M.~Kurth and R.~Sommer, {\it Renormalization and
  {O}($a$)-improvement of the static axial current},  {\em Nucl. Phys.} {\bf
  B597} (2001) 488--518, [\href{http://xxx.lanl.gov/abs/hep-lat/0007002}{{\tt
  hep-lat/0007002}}].

\bibitem{stat:actpaper}
M.~Della~Morte, A.~Shindler, and R.~Sommer, {\it On lattice actions for static
  quarks},  {\em JHEP} {\bf 08} (2005) 051,
  [\href{http://xxx.lanl.gov/abs/hep-lat/0506008}{{\tt hep-lat/0506008}}].

\bibitem{stat:letter}
{\bf ALPHA} Collaboration, M.~Della~Morte {\em et.~al.}, {\it Lattice hqet with
  exponentially improved statistical precision},  {\em Phys. Lett.} {\bf B581}
  (2004) 93--98, [\href{http://xxx.lanl.gov/abs/hep-lat/0307021}{{\tt
  hep-lat/0307021}}].

\bibitem{impr:pap1}
M.~{L\"uscher}, S.~Sint, R.~Sommer, and P.~Weisz, {\it Chiral symmetry and
  {O($a$)} improvement in lattice {QCD}},  {\em Nucl. Phys.} {\bf B478} (1996)
  365--400, [\href{http://xxx.lanl.gov/abs/hep-lat/9605038}{{\tt
  hep-lat/9605038}}].

\bibitem{hqet:pap3}
{\bf ALPHA} Collaboration, J.~Heitger, A.~{J\"uttner}, R.~Sommer, and
  J.~Wennekers, {\it Non-perturbative tests of heavy quark effective theory},
  {\em JHEP} {\bf 11} (2004) 048,
  [\href{http://xxx.lanl.gov/abs/hep-ph/0407227}{{\tt hep-ph/0407227}}].

\bibitem{nara:rainer}
R.~Sommer, {\it Non-perturbative qcd: Renormalization, o$(a)$-improvement and
  matching to heavy quark effective theory},
  \href{http://xxx.lanl.gov/abs/hep-lat/0611020}{{\tt hep-lat/0611020}}.

\bibitem{Falk:1991pz}
A.~F. Falk, B.~Grinstein, and M.~E. Luke, {\it Leading mass corrections to the
  heavy quark effective theory},  {\em Nucl. Phys.} {\bf B357} (1991) 185--207.

\bibitem{HQET:sigmabI}
G.~Amoros, M.~Beneke, and M.~Neubert, {\it Two-loop anomalous dimension of the
  chromo-magnetic moment of a heavy quark},  {\em Phys. Lett.} {\bf B401}
  (1997) 81--90, [\href{http://xxx.lanl.gov/abs/hep-ph/9701375}{{\tt
  hep-ph/9701375}}].

\bibitem{HQET:sigmabII}
A.~Czarnecki and A.~G. Grozin, {\it Hqet chromomagnetic interaction at two
  loops},  {\em Phys. Lett.} {\bf B405} (1997) 142--149,
  [\href{http://xxx.lanl.gov/abs/hep-ph/9701415}{{\tt hep-ph/9701415}}].

\bibitem{HQET:sigmabIII}
A.~Grozin, P.~Marquard, J.~Piclum, and M.~Steinhauser {\em SFB/TR~09, to be
  published}.

\bibitem{alpha:SU3}
M.~{L\"uscher}, R.~Sommer, P.~Weisz, and U.~Wolff, {\it A precise determination
  of the running coupling in the {SU(3)} {Y}ang-{M}ills theory},  {\em Nucl.
  Phys.} {\bf B413} (1994) 481--502,
  [\href{http://xxx.lanl.gov/abs/hep-lat/9309005}{{\tt hep-lat/9309005}}].

\bibitem{Eichfein}
E.~Eichten and F.~Feinberg, {\it Spin dependent forces in qcd},  {\em Phys.
  Rev.} {\bf D23} (1981) 2724.

\bibitem{Gromes}
D.~Gromes, {\it Spin dependent potentials in qcd and the correct long range
  spin orbit term},  {\em Z. Phys.} {\bf C26} (1984) 401.

\bibitem{reviews:pnrqcd}
N.~Brambilla, A.~Pineda, J.~Soto, and A.~Vairo, {\it Effective field theories
  for heavy quarkonium},  \href{http://xxx.lanl.gov/abs/hep-ph/0410047}{{\tt
  hep-ph/0410047}}.

\bibitem{pot:KK}
Y.~Koma and M.~Koma, {\it Spin-dependent potentials from lattice qcd},
  \href{http://xxx.lanl.gov/abs/hep-lat/0609078}{{\tt hep-lat/0609078}}.

\bibitem{stat:eichhill_za}
E.~Eichten and B.~Hill, {\it Renormalization of heavy - light bilinears and
  $f_b$ for {W}ilson fermions},  {\em Phys. Lett.} {\bf B240} (1990) 193.

\bibitem{SF:symmetries}
M.~Della~Morte, R.~Hoffmann, and F.~Knechtli, {\it Discrete symmetries of
  lattice qcd (and the sf)},  {\em Internal notes of the ALPHA Collaboration}
  (2005).

\bibitem{pert:1loop}
S.~Sint and R.~Sommer, {\it The running coupling from the {QCD} {Schr\"odinger}
  functional: A one loop analysis},  {\em Nucl. Phys.} {\bf B465} (1996)
  71--98, [\href{http://xxx.lanl.gov/abs/hep-lat/9508012}{{\tt
  hep-lat/9508012}}].

\bibitem{mbar:pert}
{\bf ALPHA} Collaboration, S.~Sint and P.~Weisz, {\it The running quark mass in
  the sf scheme and its two loop anomalous dimension},  {\em Nucl. Phys.} {\bf
  B545} (1999) 529, [\href{http://xxx.lanl.gov/abs/hep-lat/9808013}{{\tt
  hep-lat/9808013}}].

\bibitem{mbar:nf2}
{\bf ALPHA} Collaboration, M.~Della~Morte {\em et.~al.}, {\it Non-perturbative
  quark mass renormalization in two-flavor qcd},  {\em Nucl. Phys.} {\bf B729}
  (2005) 117--134, [\href{http://xxx.lanl.gov/abs/hep-lat/0507035}{{\tt
  hep-lat/0507035}}].

\bibitem{pot:r0_SU3}
{\bf ALPHA} Collaboration, M.~Guagnelli, R.~Sommer, and H.~Wittig, {\it
  Precision computation of a low-energy reference scale in quenched lattice
  {QCD}},  {\em Nucl. Phys.} {\bf B535} (1998) 389,
  [\href{http://xxx.lanl.gov/abs/hep-lat/9806005}{{\tt hep-lat/9806005}}].

\bibitem{pot:HMfact}
A.~Huntley and C.~Michael, {\it Spin spin and spin - orbit potentials from
  lattice gauge theory},  {\em Nucl. Phys.} {\bf B286} (1987) 211.

\bibitem{4ferm:nf0}
{\bf ALPHA} Collaboration, M.~Guagnelli, J.~Heitger, C.~Pena, S.~Sint, and
  A.~Vladikas, {\it Non-perturbative renormalization of left-left four-fermion
  operators in quenched lattice qcd},  {\em JHEP} {\bf 03} (2006) 088,
  [\href{http://xxx.lanl.gov/abs/hep-lat/0505002}{{\tt hep-lat/0505002}}].

\bibitem{4ferm:pert}
F.~Palombi, C.~Pena, and S.~Sint, {\it A perturbative study of two four-quark
  operators in finite volume renormalization schemes},  {\em JHEP} {\bf 03}
  (2006) 089, [\href{http://xxx.lanl.gov/abs/hep-lat/0505003}{{\tt
  hep-lat/0505003}}].

\bibitem{stat:zbb_pert}
F.~Palombi, M.~Papinutto, C.~Pena, and H.~Wittig, {\it A strategy for
  implementing non-perturbative renormalisation of heavy-light four-quark
  operators in the static approximation},  {\em JHEP} {\bf 08} (2006) 017,
  [\href{http://xxx.lanl.gov/abs/hep-lat/0604014}{{\tt hep-lat/0604014}}].

\bibitem{lat06:carlos}
C.~Pena, {\it Twisted mass qcd for weak matrix elements},  {\em PoS} {\bf
  LAT2006} (2006) 019, [\href{http://xxx.lanl.gov/abs/hep-lat/0610109}{{\tt
  hep-lat/0610109}}].

\bibitem{zastat:pap3}
{\bf ALPHA} Collaboration, J.~Heitger, M.~Kurth, and R.~Sommer, {\it
  Non-perturbative renormalization of the static axial current in quenched
  qcd},  {\em Nucl. Phys.} {\bf B669} (2003) 173--206,
  [\href{http://xxx.lanl.gov/abs/hep-lat/0302019}{{\tt hep-lat/0302019}}].

\bibitem{zastat:nf2}
M.~Della~Morte, P.~Fritzsch, and J.~Heitger, {\it Non-perturbative
  renormalization of the static axial current in two-flavour qcd},  {\em JHEP}
  {\bf 02} (2007) 079, [\href{http://xxx.lanl.gov/abs/hep-lat/0611036}{{\tt
  hep-lat/0611036}}].

\bibitem{thesis:skurth}
S.~Kurth, {\it The renormalised quark mass in the {Schr\"odinger} functional of
  lattice {QCD}: A one-loop calculation with a non- vanishing background
  field},  \href{http://xxx.lanl.gov/abs/hep-lat/0211011}{{\tt
  hep-lat/0211011}}.

\bibitem{pert:2loop_SU2}
R.~Narayanan and U.~Wolff, {\it Two loop computation of a running coupling in
  lattice yang- mills theory},  {\em Nucl. Phys.} {\bf B444} (1995) 425--446,
  [\href{http://xxx.lanl.gov/abs/hep-lat/9502021}{{\tt hep-lat/9502021}}].

\bibitem{pert:2loop_fin}
{\bf ALPHA} Collaboration, A.~Bode, P.~Weisz, and U.~Wolff, {\it Two loop
  computation of the {Schr\"odinger} functional in lattice {QCD}},  {\em Nucl.
  Phys.} {\bf B576} (2000) 517--539,
  [\href{http://xxx.lanl.gov/abs/Erratum-ibid.B600:453,2001,
  Erratum-ibid.B608:481,2001, hep-lat/9911018}{{\tt Erratum-ibid.B600:453,2001,
  Erratum-ibid.B608:481,2001, hep-lat/9911018}}].

\bibitem{Weisz:int_csw}
P.~Weisz, {\it Computation of the improvement coefficient $c_{sw}$ to 1-loop},
  {\em Internal notes of the ALPHA Collaboration} (1996).

\bibitem{Creutz:1984mg}
M.~Creutz, {\it Quarks, gluons and lattices}, . Cambridge, Uk: Univ. Pr. (
  1983) 169 P. ( Cambridge Monographs On Mathematical Physics).

\bibitem{deDivitiis:1995yp}
G.~M. de~Divitiis, R.~Frezzotti, M.~Guagnelli, and R.~Petronzio, {\it
  Nonperturbative determination of the running coupling constant in quenched
  su(2)},  {\em Nucl. Phys.} {\bf B433} (1995) 390--402,
  [\href{http://xxx.lanl.gov/abs/hep-lat/9407028}{{\tt hep-lat/9407028}}].

\bibitem{Cabibbo:1982zn}
N.~Cabibbo and E.~Marinari, {\it A new method for updating su(n) matrices in
  computer simulations of gauge theories},  {\em Phys. Lett.} {\bf B119} (1982)
  387--390.

\bibitem{Fabricius:1984wp}
K.~Fabricius and O.~Haan, {\it Heat bath method for the twisted eguchi-kawai
  model},  {\em Phys. Lett.} {\bf B143} (1984) 459.

\bibitem{Kennedy:1985nu}
A.~D. Kennedy and B.~J. Pendleton, {\it Improved heat bath method for monte
  carlo calculations in lattice gauge theories},  {\em Phys. Lett.} {\bf B156}
  (1985) 393--399.

\bibitem{HOR1}
M.~Creutz, {\it Overrelaxation and monte carlo simulation},  {\em Phys. Rev.}
  {\bf D36} (1987) 515.

\bibitem{PPR}
G.~Parisi, R.~Petronzio, and F.~Rapuano, {\it A measurement of the string
  tension near the continuum limit},  {\em Phys. Lett.} {\bf 128B} (1983) 418.

\bibitem{algo:LW1}
M.~L{\"u}scher and P.~Weisz, {\it Locality and exponential error reduction in
  numerical lattice gauge theory},  {\em JHEP} {\bf 09} (2001) 010,
  [\href{http://xxx.lanl.gov/abs/hep-lat/0108014}{{\tt hep-lat/0108014}}].

\bibitem{Wolff:2003sm}
{\bf ALPHA} Collaboration, U.~Wolff, {\it Monte carlo errors with less errors},
   {\em Comput. Phys. Commun.} {\bf 156} (2004) 143--153,
  [\href{http://xxx.lanl.gov/abs/hep-lat/0306017}{{\tt hep-lat/0306017}}].

\end{thebibliography}

\providecommand{\href}[2]{#2}\begingroup\raggedright
\endgroup

\end{document}